# Disclosure Notes

This manuscript is also available on both SSRN and PsyArXiv to support broader interdisciplinary access.

The authors used artificial intelligence tools, including ChatGPT, Claude, and Scite.ai, to assist in identifying relevant research literature and refining the clarity and readability of the manuscript. All final content, critical interpretation, and responsibility for accuracy remain solely with the authors.

# Corresponding Author

Barbara Oakley
Email: oakley@oakland.edu

# The Memory Paradox:
# Why Our Brains Need Knowledge in an Age of AI

Barbara Oakley,[1] Michael Johnston,[2] Ken-Zen Chen,[3] Eulho Jung,[4] Terrence Sejnowski[5]

**Abstract:** In the age of generative AI and ubiquitous digital tools, human cognition faces a structural paradox: as external aids become more capable, internal memory systems risk atrophy. Drawing on neuroscience and cognitive psychology, this paper examines how heavy reliance on AI systems and discovery-based pedagogies may impair the consolidation of declarative and procedural memory—systems essential for expertise, critical thinking, and long-term retention. We review how tools like ChatGPT and calculators can short-circuit the retrieval, error correction, and schema-building processes necessary for robust neural encoding. Notably, we highlight striking parallels between deep learning phenomena such as "grokking" and the neuroscience of overlearning and intuition. Empirical studies are discussed showing how premature reliance on AI during learning inhibits proceduralization and intuitive mastery. We argue that effective human-AI interaction depends on strong internal models—biological "schemas" and neural manifolds—that enable users to evaluate, refine, and guide AI output. The paper concludes with policy implications for education and workforce training in the age of large language models.

## 1. Introduction

Educators in recent decades have often championed "learning how to learn" and critical thinking skills over rote knowledge, encapsulated in the refrain: "Why memorize it when you can look it up?" Yet this modern mindset carries a paradox. Just as schools and students began relying on calculators and the internet, a significant shift occurred. Decades of steadily rising IQ scores—the famed Flynn Effect—suddenly leveled off and even began to reverse in several high-income countries. Although IQ is undoubtedly influenced by multiple factors, considerable evidence suggests this educational shift away from explicit content instruction and memorization, combined with increased reliance on external memory aids and continuous digital distractions, has actively contributed to declining cognitive performance. This chapter asks a provocative question: Could our increased reliance on external memory aids and digital distractions be quietly eroding the very cognitive abilities we aim to enhance?

---

[1] Barbara Oakley, Oakland University, oakley@oakland.edu
[2] Michael Johnston, New Zealand Initiative, michael.johnston@nzinitiative.org.nz
[3] Ken-Zen Chen, National Yang Ming Chiao Tung University, kenzenchen@nycu.edu.tw
[4] Eulho Jung, Uniformed Service University of the Health Sciences, eulho.jung@usuhs.edu
[5] Terrence J. Sejnowski, The Salk Institute for Biological Studies, terry@snl.salk.edu

To grapple with this question, we need to reconsider the popular educational practice that treats knowledge as separate from skills. Neuroscience reveals that knowledge and skills are not separate entities but deeply intertwined—two sides of the same coin. Recognizing this relationship helps us see why it's essential to have key information embedded in memory, rather than relying solely on external devices. This chapter explores these intriguing insights and questions, delving into the neuroscience of learning and its implications for education in the digital age.

Emerging research on learning and memory reveals that relying heavily on external aids can hinder deep understanding. Equally problematic, however, are the pedagogical approaches used in tandem with reliance on external aids—that is, constructivist, often coupled with student-centered approaches where the student is expected to discover the insights to be learned. Teachers sometimes interpret these approaches as discouraging clear, explicit explanations and corrections from the teacher. The familiar platitude advises teachers to be a *guide on the side* rather than a *sage on the stage*, but this oversimplifies reality: explicit teaching—clear, structured explanations and thoughtfully guided practice—is often essential to make progress in difficult subjects. Sometimes the *sage on the stage is* invaluable.

While humans naturally acquire certain skills like language and facial recognition without explicit teaching (known as *biologically primary knowledge*), mastering culturally important academic subjects—such as reading, mathematics, or science (*biologically secondary knowledge*)—generally requires deliberate instruction. (Sweller, 2008) Our brains simply aren't wired to effortlessly internalize this kind of secondary knowledge—in other words, formally taught academic skills and content—without deliberate practice and repeated retrieval. Excessive cognitive offloading interrupts this necessary internalization, leaving us with superficial schemata—weak mental frameworks that can't adequately support critical thinking or creative problem-solving.

At the heart of effective learning are our brain's dual memory systems: one for explicit facts and concepts we consciously recall (*declarative memory*), and another for skills and routines that become second nature (*procedural memory*). Building genuine expertise often involves moving knowledge from the declarative system to the procedural system—practicing a fact or skill until it embeds deeply in the subconscious circuits that support intuition and fluent thinking. This is why a chess master can instantly recognize strategic patterns, or a novelist effortlessly deploy a rich vocabulary—countless hours of internalizing information have reshaped their neural networks.

Internalized networks form mental structures called schemata, (the plural of "schema") which organize knowledge and facilitate complex thinking. For example, a well-developed schema in mathematics enables intuitive numerical understanding, while in literature, it enhances comprehension of narratives and themes. Schemata gradually develop through active engagement and practice, with each recall strengthening these mental frameworks. Metaphors can enrich schemata by linking unfamiliar concepts to familiar experiences (Lakoff, 2014).

However, excessive reliance on external memory aids can prevent this process. Constantly looking things up instead of internalizing them results in shallow schemata, limiting deep understanding and cross-domain thinking. This highlights the paradox: in an age saturated with external information, genuine insight still depends on robust internal knowledge

This chapter explores *why* storing knowledge in our own memory remains crucial, even (and especially) when technology offers to do the remembering for us. We delve into the science of memory to debunk the false choice between knowledge and skills, showing that a strong memory foundation actually empowers skillful thinking. We draw on neuroscientific evidence about how memory retrieval and practice strengthen learning, and we consider the cognitive consequences when people increasingly "Google it" instead of learning it. The aim is to engage both neuroscientists and educators, as well as interested readers from all backgrounds, in a conversation about the importance of *knowing* in the age of *information*. Ultimately, this is a call for balance—a vision of augmentation without atrophy. We can absolutely embrace smart technologies and abundant information, but we must also keep exercising our biological memory and attention. If we preserve that balance, we won't have to choose between a nimble mind and an encyclopedic one. In a world where we can look up almost everything, the ironic truth is that the knowledge we carry *inside our heads* is more valuable than ever.

This chapter takes a unique perspective by grounding its analysis in the latest neuroscience—illuminating the cognitive underpinnings of effective learning and the subtle ways modern educational trends may inadvertently disrupt them. As we journey deeper, we'll uncover how memory systems and schema formation shape the way we think, reason, and solve problems. Along the way, intriguing parallels between human cognition and contemporary advances in artificial intelligence will emerge, showing us both the promise and peril of integrating digital tools into our cognitive processes. Through these insights, we aim to outline a vision for an education system that cultivates minds capable of both deep understanding and effective use of digital tools.

We will also provide historical context by examining how seminal educational theorists such as Piaget, Vygotsky, Dewey, Montessori, Bruner, and Skinner anticipated—and sometimes fundamentally misunderstood—key principles of learning now clarified by neuroscience. Highlighting both their valuable insights and critical missteps helps explain how persistent misconceptions about learning emerged, even among the most influential educational figures. Their legacies, while groundbreaking, were often double-edged, simultaneously driving innovation and inadvertently creating barriers to more effective educational practices.

**Glossary: Key Concepts in Learning and Memory**

**Cognitive Offloading:** Reducing cognitive load by relying on external tools (such as smartphones, calculators, or internet searches) to store or handle information. While this reduces immediate cognitive effort, excessive offloading can weaken internal memory formation, limiting the development of robust, flexible memory networks.

**Declarative Memory System:** The brain's system for consciously recalling facts, concepts, and experiences. Primarily involving the hippocampus and medial temporal lobe, this system allows explicit, deliberate retrieval of information.

**Engram:** The physical trace a memory leaves in your brain, created when groups of neurons strengthen their connections during learning or experience. Engrams physically store your memories and integrate new information into your existing mental frameworks—your schemata.

**Interleaving:** A learning technique in which practice sessions alternate among related but distinct topics or skills. Interleaving challenges learners to repeatedly distinguish between different memory traces (engrams), strengthening each memory and enhancing the ability to transfer knowledge across contexts.

**Neural Manifolds:** Patterns of coordinated neural activity among groups of neurons, allowing the brain to efficiently simplify and store complex information. Schemata often correspond to specific neural manifolds, reflecting stable, organized neural activity patterns developed through experience. However, not all manifolds represent schemata; some capture simpler sensory or motor activities without supporting higher-level organization. Unlike an engram – the physical trace a memory leaves in the brain's neural connections – a neural manifold is the dynamic pattern of neural firing that represents or recalls that information.

**Procedural Memory System:** The brain's system for gradually acquiring habits, skills, and routines through repeated practice. Centered mainly in the basal ganglia and related cortical regions, procedural memory supports automatic performance of tasks, often without conscious awareness.

**Long-Term Working Memory:**
A recently proposed form of memory that bridges the gap between short-term working memory (lasting seconds) and long-term memory (lasting days to years). Long-term working memory allows the brain to hold ideas active for minutes or even hours—long enough to integrate and reflect on concepts across extended learning sessions. It is thought to be supported by temporary but longer-lasting changes in synaptic strength, especially through mechanisms like spike-timing-dependent plasticity (STDP). This memory system helps explain how learners can make meaningful connections between ideas presented far apart in time—such as linking concepts from the beginning and end of a long lecture—before those ideas are fully consolidated into long-term memory. Long-term working memory plays a crucial role in real-time understanding, reasoning, and question generation, and helps illustrate why internal memory—even if transient—is essential for deep learning

> **Retrieval Practice:** Actively recalling information from memory rather than passively reviewing it. Retrieval practice strengthens memory by repeatedly reactivating and reinforcing neural connections (engrams), enhancing long-term retention. Interleaving complements retrieval practice by increasing the challenge of differentiating among related concepts, further solidifying memories.
>
> **Schemata:** Mental frameworks that organize knowledge into meaningful patterns. (*Schema* is singular and *schemata* is plural.) Schemata contain variables that adapt to different situations and nest hierarchically (smaller schemata within larger ones). Schemata actively generate expectations based on prior knowledge, enabling rapid understanding of new information. Unlike engrams (physical neural traces), schemata are abstract structures that could exist outside the brain—in computer code or written text, for example.
>
> **Spaced Repetition/Spaced Retrieval/Spaced Learning:** A method of learning that spreads practice sessions out over time, instead of cramming them together. Neuroscience shows this spacing prompts the brain to repeatedly revisit and reinforce the material, making memories clearer, stronger, and longer-lasting.

## 2. The Memory Traces: Understanding Engrams

### 2.1 The Physical Nature of Memory

An engram is the physical trace a memory leaves in your brain. When you learn something new or have an experience, groups of neurons activate together, forming stronger connections. These strengthened neural connections form the engram—the biological foundation of memory.

But memories aren't stored like exact files on a computer. Instead, your brain connects each new memory to what you already know, integrating it into your existing mental frameworks—your schemata. Schemata are abstract structures that organize your experiences. For instance, your schema of a "dog" combines common details (four-legged, furry, barks) from many past encounters.

When you remember something, you don't simply replay an exact recording of the event. Instead, you reconstruct the memory, filling in missing details using your schemata. This is why your memories are flexible, efficient, and occasionally prone to errors. Put simply, schemata represent what you know, and engrams represent how your brain physically stores and retrieves that knowledge. While schemata can exist in many forms—including computer code or written text—the term "engram" specifically refers to biological memory traces in your brain. (Guskjolen and Cembrowski, 2023)

Think of an engram as a neural "imprint" left by what you've learned. When you commit something to memory—whether it's a mathematical fact like $4 \times 7 = 28$ or a vivid personal event—specific neurons fire together, strengthening their connections and creating a distinctive network. This network becomes the physical trace of that memory, lasting long after initial learning.

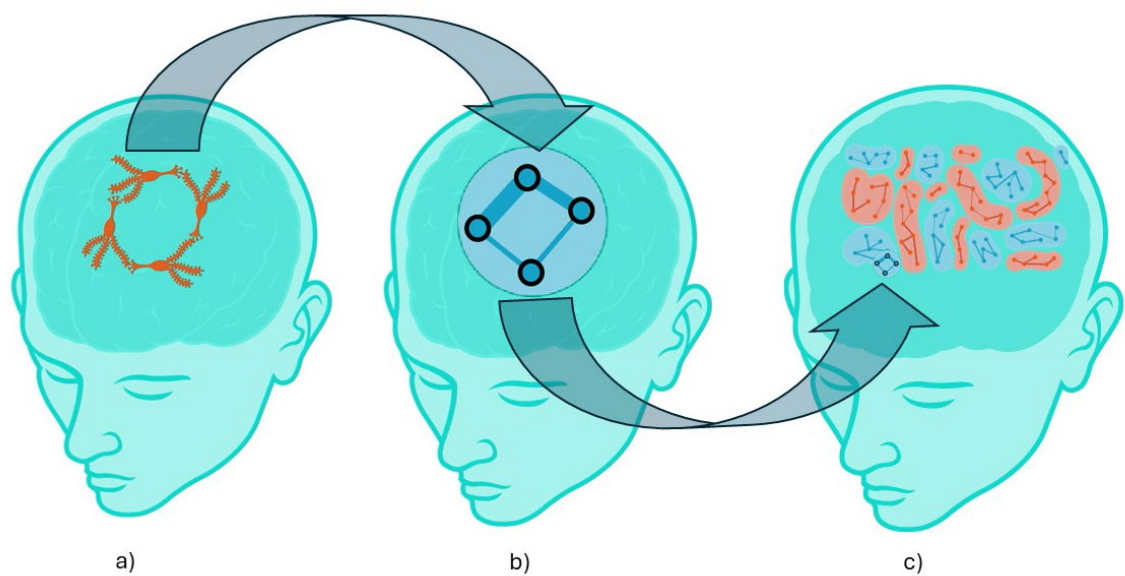

**Figure 1: Engrams and schemata a)** A simplistic representation of the connected neurons of a memory (an "engram"). **b)** The connected cluster of neurons of an engram is typically shown by dots and lines. Although only four "neurons" are shown here, this engram representation actually represents clusters of neurons scattered around the brain which activate together when we retrieve that engram from memory. **c)** Schemata organize memory by providing abstract frameworks that can activate relevant engrams. Here, we show declarative engrams in blue and procedural engrams in orange, both organized within larger schema structures. Unlike engrams (physical neural traces), schemata represent the patterns of organization themselves. More about this in Section 3.

## 2.2 Encoding and Retrieval Mechanisms

The way engrams function reveals much about how memory works. Research confirms the *encoding specificity principle*—memory retrieval works best when current conditions match those present during learning. This explains why returning to a place where you first learned something can suddenly trigger vivid recollections; the environmental cues reactivate the associated engram. A related phenomenon is context-dependent learning: if you encode information while drunk, for example, you can recall it better when drunk than when sober. This phenomenon demonstrates how context affects memory retrieval—though it's not a strategy educators would recommend adopting.

During successful memory recall, the same neurons that fired during the original learning experience become active again. What's striking is that activating just a small subset of these engram neurons can trigger the entire memory network through "pattern completion." This neural cascade effect allows complete memory retrieval from partial cues, demonstrating the interconnected nature of memory representations.

But note that a cue about where to find a memory is quite different than a cue that activates the memory itself. For instance, a student remembering that they can ask an AI to explain photosynthesis doesn't activate the same neural networks as actually remembering how plants convert sunlight into energy. This distinction between memory pointers and actual knowledge represents a core theme of this chapter that we'll return to: the difference between knowing *where to find* information versus *truly incorporating that information into our internal schemata and neural architecture.*

The brain also employs a performance-enhancing mechanism: when you successfully retrieve a memory, the engram neurons temporarily become more excitable. This heightened state can last for hours, making subsequent retrievals easier and more accurate. This post-retrieval excitability directly affects memory performance and provides a neural basis for why spaced retrieval is so effective. (Pignatelli *et al.*, 2019; Carpenter, Pan and Butler, 2022)

**2.3 Memory Consolidation Processes**

This brings us to the process of memory consolidation—how memories stabilize over time. (See Figure 2.) Consolidation involves two distinct but related processes occurring at different levels. Synaptic consolidation refers to the post-encoding transformation of information into a long-term form at local synaptic and cellular nodes in the neural circuit that encodes the memory. During this process, connections between neurons are both strengthened and pruned, creating a more efficient memory trace. Synaptic consolidation is traditionally assumed to draw to a close within hours of its initiation. Systems consolidation, on the other hand, refers to the post-encoding reorganization of memory representations over distributed brain circuits. It involves recurrent waves of synaptic consolidation in new brain regions as the memory representation gradually spreads beyond its original location. During sleep, particularly slow-wave sleep (SWS), the hippocampus 'replays' memory activity patterns, facilitating the gradual transfer of memories to distributed cortical networks. Systems consolidation may last days to months and even years, depending on the memory system and the task. (Dudai, Karni and Born, 2015)

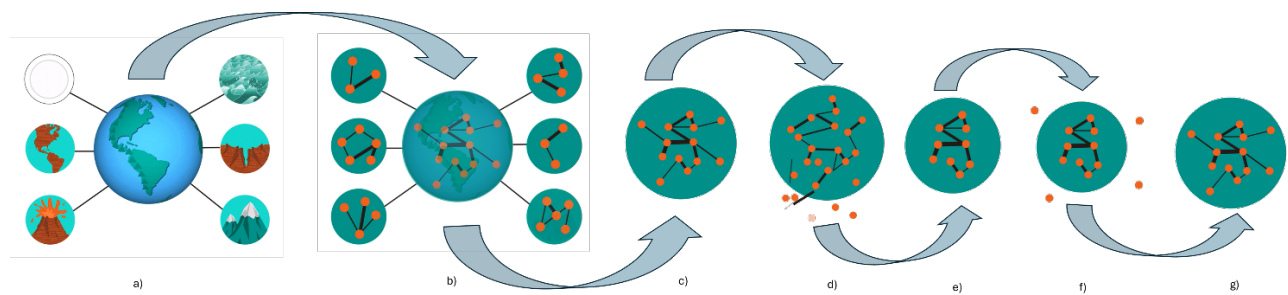

**Figure 2: Memory consolidation.** a) When learning a new concept, for example, like plate tectonics, there are extraneous bits of information, like "plates" have food and mountains can be cold. (b) and (c) The first links that come together have extraneous neural clutter. (d) As the concept is used, the extraneous links fall away (e) to leave the elegant gist of the concept. (f) and (g) As consolidation processes firm into place, links to other concepts can form, like the physics of friction, mathematical rates of movement, or social shifts. These interconnections not only enrich understanding but also create cognitive scaffolding that facilitates faster learning of related concepts and more creative problem-solving in novel situations. This process exemplifies how schemata become tuned and integrated into larger knowledge networks.

Recent research reveals a selective mechanism for memory consolidation. Our brains don't preserve all experiences equally—they actively select which memories to strengthen. During moments when we encounter novelty or surprise, process rewards, or complete learning tasks, the hippocampus generates sharp wave ripples (SPW-Rs) that replay and tag significant experiences. These tagged experiences are then preferentially replayed during sleep, creating a direct pathway from rewarding experience to lasting memory. (Yang et al., 2024)

This mechanism has been directly observed in studies with mice. During momentary pauses in exploration—particularly at reward locations—the hippocampus shifts from theta rhythms to generating sharp wave ripples (SPW-Rs) that replay recent experiences. Similar ripple activity occurs in the human hippocampus during periods of quiet wakefulness and rest. This suggests that moments of reflection after completing a task or receiving feedback may enhance memory through similar neural tagging mechanisms, influencing which experiences our brains select for long-term storage.

For humans, this has important implications for learning practices. When we immediately reach for our phones after completing a learning task or solving a problem, we may be disrupting this critical neural tagging process. Instead, brief periods of undistracted reflection after learning could provide a neurophysiological window when memories are specially marked for preservation. This may explain why information processed during focused attention tends to be better remembered than information encountered during distracted states, and suggests that intentional pauses for reflection—rather than digital distraction—could significantly enhance our ability to consolidate important experiences into lasting memories.

When you actively retrieve a memory, you're not just accessing it, but rather, are actively reinforcing the engram by engaging the entire neural network. This strengthens connections throughout the engram and begins linking to other concepts—a process far more helpful for learning than simply re-reading or passively reviewing the material. (Wamsley, 2019)

As memories consolidate, they can undergo important transitions in how they're stored and accessed in the brain. With repeated use and retrieval practice, initially consciously-recalled information can become more automatic and intuitive. This often involves a transition between the brain's two memory systems, and is crucial for developing expertise.

# 3. Declarative vs. Procedural Memory: Two Learning Systems

### 3.1 Understanding the Two Memory Systems

The human brain has two major learning and memory systems that work in parallel: one for facts and events (*declarative memory*) and one for skills and habits (*procedural memory*). (Figure 3) These rely on distinct neural circuits—roughly speaking, the hippocampus and related medial temporal lobe structures support declarative memory, whereas the basal ganglia[1] (especially the striatum) and frontal cortex support procedural memory. From an evolutionary perspective, the declarative system, which is accessible to our conscious thoughts, is believed to have developed more recently, while the procedural system, which we're not conscious of, is much older. (Ten Berge and Van Hezewijk, 1999) The basal ganglia's structures, in fact, appeared early in vertebrate evolution and have been highly conserved, suggesting they serve essential functions.

---

[1] The cerebellum and other structures are involved as well, but let's not go there in this discussion.

A rough way to feel the difference between these two systems is to compare your conscious, step-by-step working through a math problem (*declarative*) with your repeated attempts to learn to hit a baseball (*procedural*). In this latter case, you are only conscious of whether you hit the ball, but *not* how your procedural system is learning to hit the ball. Understanding the difference between the declarative and procedural systems is key to appreciating why certain knowledge (like multiplication tables or vocabulary) needs to be stored in the brain through practice, rather than constantly outsourced, for genuine intuition and fluency to develop. (Morgan-Short and Ullman, 2022)

### 3.2 The Declarative Learning Pathway

Declarative memory, as we had mentioned, is the system for consciously accessible knowledge—the things we can deliberately recall and state. It includes semantic memory (general knowledge and facts) and episodic memory (personal experiences). Compared to the procedural system, the declarative system learns information relatively rapidly. For example, you can hear a historical fact or a definition a single time and potentially remember it. The trade-off for this relatively quick learning is the counterintuitive fact that declarative recall can be slow and effortful than procedural recall. Learning something new through the declarative system depends critically on the hippocampus. This structure helps bind together the elements of new information and consolidate it into long-term memory. If the hippocampal system is damaged, as when a person develops Alzheimer's, it becomes difficult or impossible to learn new information.

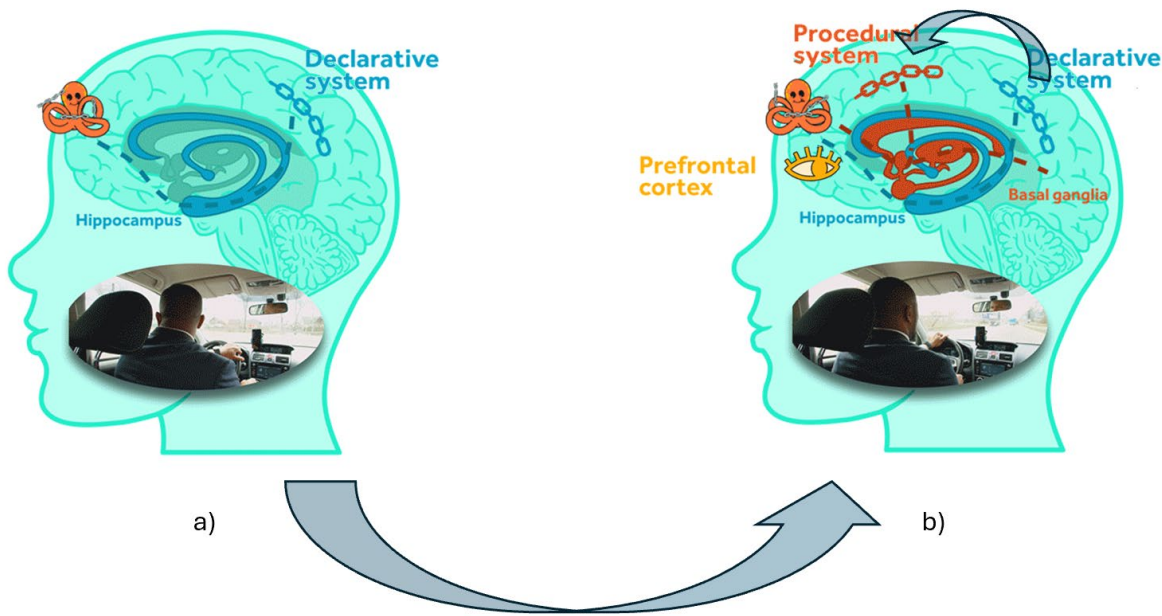

**Figure 3: The Declarative and Procedural Learning Systems.** (a) The declarative pathway takes information from working memory (symbolized by the four-armed octopus), and moves it into links in long-term memory through the hippocampus. (Cowan *et al.*, 2020) When you might first learn the route to take to drive home, you are using your declarative learning pathways as you focus on following the directions. (b) As you repeatedly drive home over the same route, a part of the prefrontal cortex toward the front of the brain is "watching." If your prefrontal cortex sees that you're driving the same route home day after day, it eventually creates procedural, habitual links in long term memory.

An important feature of declarative memory is that it's flexible: you can apply a fact learned in one context to a different context, or combine bits of knowledge in novel ways. This flexibility underpins complex reasoning and problem-solving, but it also means declarative recall tends to require conscious attention. For instance, if you've learned a formula but not practiced it much, you must deliberately recall each step to apply it. Explicit instruction—for example, a typical "talk and chalk" lecture, feeds information to students through their declarative pathways.

### 3.3 The Procedural Learning Pathway

Procedural memory, by contrast, is the system for acquiring sequences and routines that operate at an intuitive, automatic level. Riding a bicycle, playing a musical instrument, speaking grammatically in your native language, effortlessly recalling a well-practiced phrase like "Yo hablo Español," or performing mental arithmetic with ease are all supported by procedural learning. It's generally hard for us to verbalize this kind of implicit learning—you can't easily explain the muscle movements of riding a bike, the finger movements of a piano scale, or even the mental steps to reduce a fraction.

Procedural learning is typically slower to acquire and requires repetition and deliberate practice—but once learned, the skills can be executed rapidly and with little conscious effort. In the brain, procedural memory is rooted in circuits linking the cortex, basal ganglia, and cerebellum. As control shifts to more automated motor or premotor circuits, the task becomes ingrained "in the fingers" or in whatever neural substrates handle the routine. At that stage, performing the skill, task, or thought feels almost effortless—one might term it "second nature."

> **From Autopilot to Insight: Reimagining the Basal Ganglia**
>
> For decades, neuroscientists viewed the basal ganglia simply as the brain's center for habitual learning—helping us ride bikes or type without conscious thought. Recent research reveals a far richer picture.
>
> Neuroimaging now shows these neural structures underpin our ability to recognize complex patterns across diverse domains. In language, the basal ganglia activate during intricate grammatical parsing, regardless of whether language is spoken or signed. Activity increases with linguistic complexity, suggesting these regions actively decode and predict grammatical patterns. (Moreno *et al.*, 2018)
>
> Similarly, in mathematics, the basal ganglia extend beyond storing arithmetic facts to recognizing mathematical patterns that enable intuitive leaps. Patients with basal ganglia impairments struggle specifically with multi-step procedures, highlighting these structures' role in intuitively applying mathematical concepts rather than merely recalling isolated facts. (Saban *et al.*, 2024)
>
> This sophisticated functionality extends to general cognition as well. Structures like the caudate nucleus activate strongly during moments of sudden insight—those "Aha!" experiences that mark creative breakthroughs. This suggests the basal ganglia help us break from habitual thinking to form novel connections between ideas. (Tik *et al.*, 2018)
>
> The basal ganglia serve not merely as repositories for practiced routines but as essential components in our ability to detect, learn from, and intuitively apply complex patterns. This expanded understanding helps explain how deep expertise and fluent intuition emerge from what initially appears to be simple repetitive practice.
>
> Interestingly, such deep intuitive learning parallels a phenomenon in artificial intelligence called "grokking." In grokking, AI models seem initially stuck during repetitive training, only to suddenly exhibit dramatic improvements in their ability to generalize—long after mastering the training data. Previously, this extended training was viewed negatively as "overfitting" or "overtraining," similar to what educators have traditionally criticized as "overlearning." However, grokking reveals that repetitive practice may quietly set the stage for profound cognitive breakthroughs, transforming habitual routines into deeper intuitive understanding. (Power *et al.*, 2022)

**3.4 Complementary Systems: How They Work Together**

Why do we have these two systems? They complement and sometimes even duplicate one another—providing richer ways for the brain to work with information.

With repeated use, memories evolve within the brain's architecture. New declarative memories—facts, events, and explicit knowledge like vocabulary words, historical dates, or scientific formulas—begin their life dependent on the hippocampus, with all its rich contextual detail. In contrast, procedural memories, like learning to kick a soccer ball or type without looking at the keyboard, initially form through neural pathways largely involving the basal ganglia.

Recent neuroscience research clarifies why spaced learning is so powerful in this process. When you first learn something declaratively, your hippocampus rapidly encodes a temporary explicit memory by linking together neurons across the cerebral cortex. As we retrieve these declarative memories repeatedly over spaced intervals, something transformative happens: these cortical connections strengthen, enabling the memories to eventually become stable and independent of the hippocampus. This is the critical transition point—these stabilized cortical memories now serve as clear blueprints that allow deeper brain structures—especially the basal ganglia—to gradually automate the knowledge. The hippocampus gradually releases its hold as the basal ganglia take greater responsibility for these memories. This shift has been directly observed—as memories mature, activity decreases in the hippocampus while increasing in the striatum, a key structure of the basal ganglia. After this proceduralization by the basal ganglia, knowledge that was once explicit and deliberate becomes implicit, automatic, and intuitive. (Goldfarb, Chun and Phelps, 2016; Narasimhalu et al., 2012; Buch et al., 2021; Thompson et al., 2024; Yang et al., 2025)

Declarative memory is like a fast-learning but somewhat slow-operating notebook—flexible, but not optimized for speed during retrieval. Procedural memory is like a subroutine that, once trained, runs blazingly fast and automatically. Both physical skills (like playing an instrument) and mental tasks (like solving equations) follow this pattern: practice transforms explicit declarative knowledge into automatic procedural memory while preserving the original declarative engram. This dual access allows us to think flexibly while acting intuitively. (Packard and Knowlton, 2002)

Practically, this means that in many learning domains, *both* systems are involved. This interplay between systems explains the artificial divide between "knowledge" and "skills" we mentioned earlier. Take multiplication as an example: when a child is learning the multiplication table, at first she may rely on declarative memory (memorizing that 4 x 3 = 12). With repeated practice, the relationship 4 x 3 = 12 becomes a quick rote response—it transforms into a procedural memory, something she *just knows* without having to think about it. If she might see $\frac{12}{3}$, she can much more easily "feel" that it is equivalent to 4. If she never memorizes the multiplication tables and instead presses 4 x 3 on her calculator, externalizing the operation each time she does it, she develops no internal feel for the relationships between 4, 3, and 12. Even simply trying to reduce a fraction leaves her stuck with slower, conscious, declarative processing, which burdens her working memory during complex tasks.

The difference between declarative versus procedural recall of information is one we all experience: the struggle of recalling newly learned information compared to the effortless way we access deeply familiar knowledge. This process feels different because it is different—access has largely moved from a system requiring conscious effort to one that works with automatic precision. Our most treasured knowledge thus migrates from conscious deliberation to intuitive understanding—a neural efficiency that serves us well in tasks requiring speed and fluency. A challenge in learning is that when children are young, the procedural system is the stronger of the two—it's what allows us to pick up the intuitive patterns of our native language.  But as children mature, their working memory capacities increase and the declarative system comes to the fore. Thus, methods effective for teaching young children — such as hands-on manipulatives and learning through experience — do not work as well for older students who need more explicit instruction to master biologically secondary academic material. In other words, as children mature and tackle advanced subjects, they require clearer guidance and practice to learn efficiently.

**3.5 The Consequences of Offloading**

What happens if we shortcut the shift from declarative to procedural memory-making by instead using an external device? Simply put, if a learner leans too heavily on external aids, the "proceduralization" of knowledge may never fully occur. A student who always uses a calculator for basic arithmetic, for instance, might pass tests, but she may not develop the same number sense and intuition as one who has internalized those operations. The latter student, having memorized math facts and practiced operations, can *recognize patterns* (e.g. spotting that 12 x 100 = 1200) and estimate or manipulate numbers with confidence. The former student might solve every problem by brute force lookup or computation, never quite forming an intuitive grasp—the kind of intuition that often leads to creative problem-solving or spotting errors at a glance.

Without practice, the basal ganglia circuits don't get the training trials needed to optimize the skill. The result is a learner who remains reliant on conscious, declarative processing (or worse, on the tool itself) for tasks that could have become automatic. This not only makes the tasks slower; it also chokes off higher-order cognition, because the person's working memory is consumed by basics. Cognitive load theory tells us that when lower-level elements of a task are automated, working memory is freed to focus on complex aspects. For example, solving a physics problem is immensely easier if the algebra steps are second-nature; one can devote thought to the physics concepts instead of getting lost in algebraic manipulations.

Recent neuroscience research has shown that under specific conditions, procedural memory engrams can form rapidly—within hours rather than weeks or months. This can happen when new information connects to deeply internalized prior knowledge (as opposed to knowledge arising from just knowing where to find information). Prior knowledge helps memories stabilize quickly in cortical networks with less hippocampal involvement—an insight that underscores the distinction between internally stored knowledge and mere awareness of external information sources. (Hebscher et al., 2019)

Recent research also suggests that the brain may use a kind of "synaptic handwriting"—temporary but longer-lasting changes in the strength of connections between neurons—to hold ideas in mind for minutes or even hours. This mechanism, known as spike-timing-dependent plasticity, was once thought to support only long-term memory. But new findings suggest it may also underlie what some researchers call long-term working memory—a bridge between short-term recall and lasting storage that helps us keep information mentally available during extended periods of thought. This may explain how a student, near the end of a 90-minute lecture, can ask a thoughtful question that links ideas introduced much earlier in the session—even though none of those ideas have yet been committed to long-term memory. It's a powerful reminder that internal memory—even if fleeting—is what allows us to make real-time connections and insights. In contrast, simply looking something up may retrieve a fact, but it doesn't create this temporary mental workspace or support the flexible integration that deep learning depends on. (Sejnowski, 2025)

In short, internalizing foundational knowledge—both facts and procedures—is what allows the mind to synthesize and create. Over-reliance on external memory can leave one with a collection of correct outputs (answers obtained from tools) but without the integrated *understanding* or procedural fluency that marks true expertise. As the brain's two learning systems demonstrate, deep learning is a matter of training the brain as much as informing the brain. If we neglect that training by continually outsourcing, we risk shallow competence. The next section will delve more into how the brain organizes learned knowledge into efficient structures (schemata and neural representations), and how failing to internalize knowledge may disrupt these structures.

**Learning by Reward: How Reinforcement Principles Illuminate Human Memory Formation**

Computer scientists Richard Sutton and Andrew Barto revolutionized our understanding of learning through their reinforcement learning framework. Though originally developed for artificial intelligence, their work directly models how human brains learn through experience and feedback. Research has confirmed that human learning employs remarkably similar reinforcement mechanisms, with neural circuits that evaluate outcomes and strengthen successful behaviors. This evidence contradicts the popular but misguided educational notion that unguided discovery learning is optimal for academic subjects, and instead demonstrates why structured, repeated practice with clear feedback is essential for strengthening neural pathways. (Sutton and Barto, 2018)

The central insight that applies directly to human memory formation is the concept of prediction error—the gap between what we expect and what actually happens. In reinforcement learning algorithms, this prediction error signal drives learning by updating the system's knowledge. Remarkably, neuroscience discovered that the brain's dopamine system operates on precisely this principle. When something good happens unexpectedly, dopamine neurons fire briskly; when an expected reward fails to appear, they pause. This dopamine response creates what neuroscientists call "eligibility traces"—temporary tags in neural pathways that mark connections for strengthening, essentially implementing nature's version of reinforcement learning algorithms. (Sutton, 1988; Nasser et al., 2017; Shouval and Kirkwood, 2025)

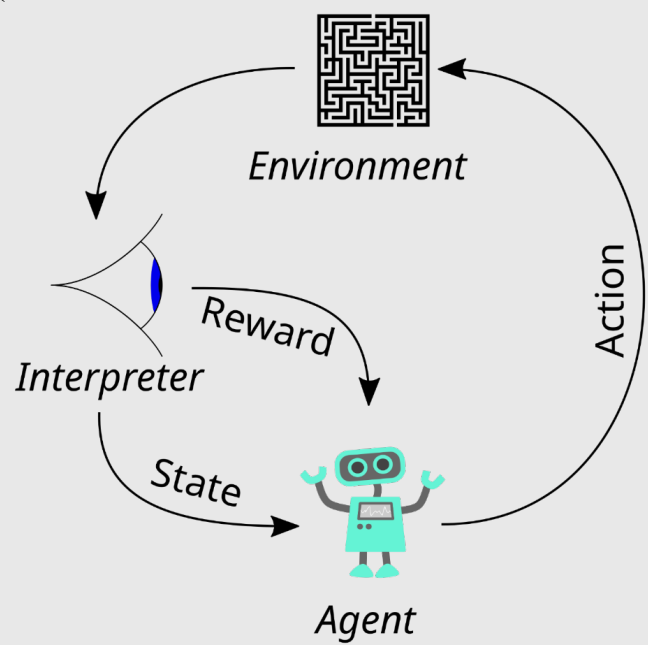

**Figure 4:** In reinforcement learning, an agent learns by taking actions and receiving feedback, either rewards or penalties. Learning occurs when outcomes differ from predictions—a process known as prediction error. The human brain also learns this way: positive prediction errors release dopamine, reinforcing successful actions. Negative prediction errors—like the single painful event of touching a hot stove ("Ouch! That stove is hot!")—rapidly teach us what to avoid in the future. This immediate, single-experience learning is similar to what's known as "one-shot learning" in artificial intelligence. (Public domain, 2017)

> This framework directly illuminates how our brains form and refine schemata—the mental frameworks discussed throughout this chapter. As we practice a task under structured conditions with appropriate feedback, reward-driven mechanisms gradually build internal models of what works. (Bein and Niv, 2025) Over time, the brain compresses these experiences into streamlined representations, just as AI systems distill countless trials into efficient response patterns.
>
> In neuroscience, these efficient representations are called neural manifolds—patterns of neural activity that capture the essence of repeated behaviors or concepts while dramatically reducing complexity. Think of them as the brain's way of creating simplified "neural shadows" that preserve essential relationships while filtering out enormous amounts of noise. Through reinforcement-like feedback, our brains sculpt these manifolds so that common situations can be recognized and responded to with minimal effort. Similarly, artificial systems trained with reinforcement learning principles compress extensive training experiences into compact computational representations, allowing for rapid responses without exhaustive processing. In both cases, learning optimizes for efficiency: redundant details are pruned away and useful patterns are reinforced. (Pao *et al.*, 2021)
>
> This convergence between human memory and artificial learning systems reveals a profound insight: prediction errors drive learning across both domains. This parallel explains why knowledge internalized in memory enables far more efficient learning than constant cognitive offloading to external devices. When we consistently outsource thinking to technology, we effectively bypass our brain's natural reinforcement learning mechanisms—the very pathways that build and strengthen neural manifolds supporting sophisticated thinking. These findings also contradict popular notions of unstructured discovery learning, demonstrating instead that developing expertise requires structured practice with clear, timely feedback that allows our reinforcement mechanisms to optimize neural representations. (Sejnowski, 2024)

# 4. The Important Role of Prediction Errors

### 4.1 Prediction Errors: The Brain's Learning Signal

Imagine you're quickly typing a familiar calculation into a calculator—perhaps you're multiplying 5 × 10. Instead of the expected "50," the calculator suddenly displays "500." Instantly, you sense something is wrong. (Figure 5) This immediate feeling of surprise isn't just confusion; it's your brain's way of signaling a prediction error—the mismatch between what your brain expected and what actually happened.

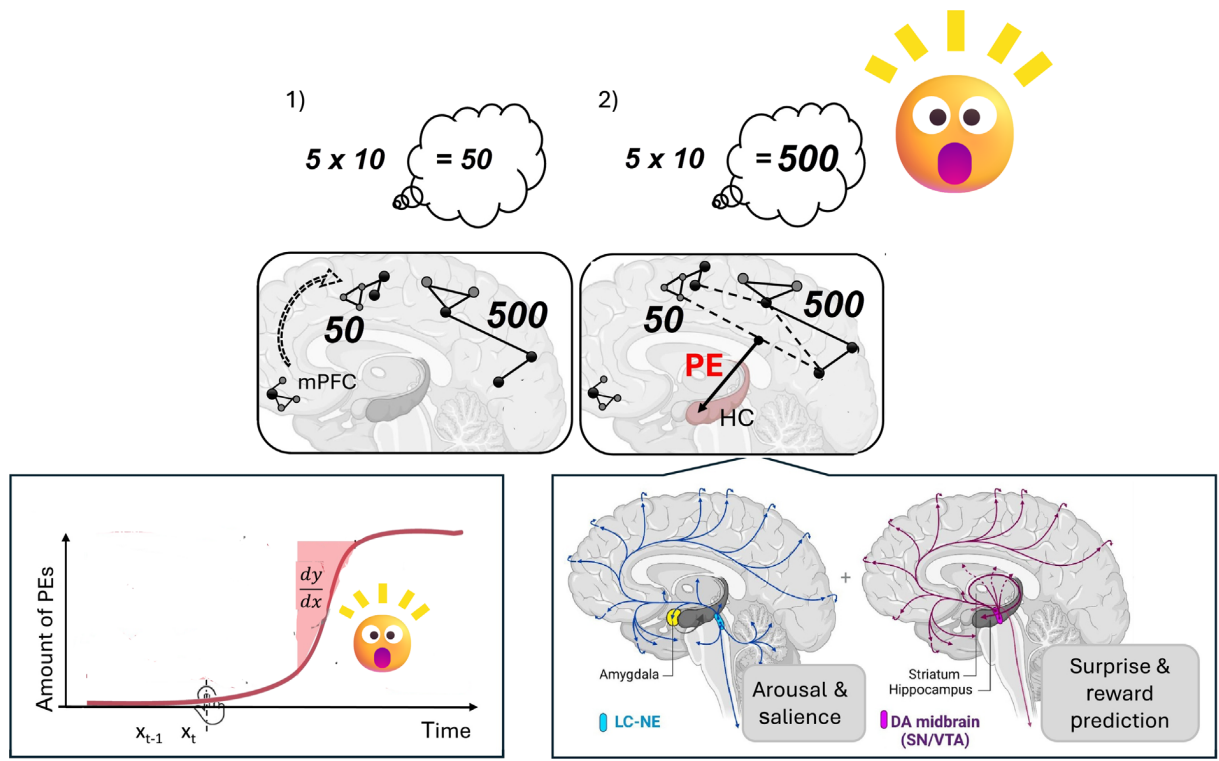

**Figure 5:** The neural basis of prediction errors. When an expected outcome differs from the actual result, dopamine neurons release neurotransmitters that tag recently active neural connections. These tags, called "eligibility traces," mark specific connections for strengthening, allowing the brain to update its internal models and form new memories based on the error. (Adapted from Becker and Cabeza, 2025).

## 4.2 The Neuroscience of Surprise

At a deeper neurological level, this feeling of surprise is orchestrated by neurotransmitters like dopamine and norepinephrine. Dopamine neurons, especially in areas like the ventral tegmental area, act as the brain's internal learning signal. They spring into action whenever outcomes don't align with our internal predictions, creating what neuroscientists call "eligibility traces"—temporary tags in active neural connections that mark them for change—either reinforcing or weakening. (This echoes Jean Piaget's insight from the early 1900s regarding *cognitive disequilibrium*.[1]) Meanwhile, norepinephrine serves as a kind of mental alarm bell, quickly enhancing alertness and sharpening focus. Together, these processes help your brain recognize, process, and learn from errors.

These fundamental learning mechanisms work not just for dramatic insights but for all feedback-based learning, from formal education to Kahoot! (flashcard) exercises. (This validates B. F. Skinner's emphasis on reinforcement and feedback in learning.) Negative experiences also powerfully shape learning. Even a single painful or unpleasant event can quickly teach us to avoid making the same mistake again—as with the hot stove example mentioned in Figure 4. (Wimmer and Büchel, 2021)

---

[1] Cognitive disequilibrium includes: *Assimilation*: fitting new information into existing mental frameworks (e.g., kids calling a penguin a bird), and *accommodation*: requires modifying those frameworks when information doesn't fit (e.g., realizing penguins can't fly)—concepts Piaget (1936) pioneered that remain fundamental to learning theories, though he didn't use "schema" in its modern sense.

### 4.3 The Necessity of Internal Knowledge

Yet there's a crucial catch: this powerful error-detection system depends on having strong internal expectations—built through memorization and repeated practice. Consider multiplication tables: a nursing student who memorized these facts as a youngster has created her own internal "schema," mental frameworks that reliably predict correct answers. When an incorrect answer pops up, her brain instantly notices something is amiss, activating the alert system and facilitating immediate correction and learning.

By contrast, students who rely entirely on calculators never fully build these strong internal schemata. When confronted with the same incorrect result, there's no mismatch, because without knowledge of the multiplication tables, there's no prediction being made. These students might accept the wrong answer at face value without realizing it's an error. The result? A nurse with no internalized multiplication tables can enter a wrong number in the calculator, and then be oblivious to the fact that the resulting medication she is dispensing is an order of magnitude larger than it should have been.

### 4.4 Prediction Errors and Learning Mechanisms

Without internally stored knowledge, our brain's natural learning mechanisms remain largely unused. Every effective learning technique—whether retrieval practice, spaced repetition, or deliberate practice—works precisely because it engages this prediction-error system. When we outsource memory to devices rather than building internal knowledge, we're not just changing where information is stored; we're bypassing the very neural mechanisms that evolved to help us learn.

In short, internalized knowledge creates the mental frameworks our brains need to spot mistakes quickly and learn from them effectively. These error signals do double-duty: they not only help us correct mistakes but also train our attention toward what's important in different contexts, helping build the schemata we need for quick thinking. Each prediction error, each moment of surprise, thus becomes an opportunity for cognitive growth—but only if our minds are equipped with clear expectations formed through practice and memorization. (Bein and Niv, 2025)

So, what is a schema?

# 5. Schemata, Neural Manifolds, and Cognitive Efficiency

### 5.1 Foundational Concepts: Schemata and Mental Frameworks

The brain doesn't store knowledge as isolated facts but organizes related information into schemata—active computational structures that both represent knowledge and contain information about how that knowledge should be used. A schema is not merely a passive framework but a recognition device that evaluates its "goodness of fit" to incoming data. While engrams are the physical imprints of memory in neural tissue, schemata are the abstract knowledge structures they represent. In other words, schemata are what we know, while engrams are how our brains physically store this knowledge. A schema is the organized pattern of information itself—one that could, in principle, exist in any medium, from neurons to computer code.

Schemata have several key characteristics: they contain variables with constraints that define typical values and relationships; they can embed within one another hierarchically; and they function across all levels of abstraction, from sensory patterns to abstract concepts. When we encounter new information, schemata not only help us interpret it but also generate expectations about unobserved aspects—providing default values that guide inference and prediction.

Cowan (2014) describes working memory as the "cauldron" of schema formation, where new concepts are born by simultaneously holding and binding together multiple elements of information. Experiments demonstrate that information must be simultaneously present in working memory for meaningful associations—and thus schemata—to form. If too much information is presented or externalized so that it cannot be concurrently maintained in working memory, meaningful connections and robust schema formation are impaired.

You can think of schemata as similar to organizing shelves in a closet, grouping related items neatly together. A schema for "restaurant," for example, includes a sequence of events (enter, get seated, read menu, order, eat, pay, exit) and expectations about roles (waiter, customer, chef) and objects. Schemata allow us to encode and retrieve information efficiently: instead of remembering every tiny detail as separate bits, we remember a unified structure with slots that can be filled in. New information that fits a schema is learned more easily, and remembering one part of a schema can cue the rest. Recent research confirms that activating schemata actively shapes memory, attention, and how we segment experiences into meaningful events. (De Soares *et al.*, 2024; Wickelgren, 2025)

In problem-solving, invoking the right schema can guide you to a solution by analogy or pattern recognition. Importantly, schemata are not innate—they are learned and refined with experience, which means they depend on memory of past instances. As schemata become richer, they allow quicker learning and better recall of new information. The formation of schemata is a hallmark of expertise. A master chess player, for instance, has schemata for typical piece configurations and strategies, allowing them to remember positions and foresee outcomes far better than a novice (the novice lacks those organized patterns in memory).

### 5.2 Neural Basis of Knowledge Organization

Immanuel Kant (1724–1804) introduced the concept of *schemata* in *Critique of Pure Reason* (1781) as part of his effort to reconcile two opposing philosophical traditions: *empiricism*, which holds that knowledge arises from sensory experience, and *rationalism,* which claims that some knowledge originates in the mind itself. Kant argued that while we receive information through the senses, we also use built-in mental categories—such as space, time, and causality—to organize it. *Schemata*, in his account, are mental procedures that allow these abstract categories to be applied to concrete experience—helping us, for example, perceive events not just as isolated flashes, but as unfolding in time and governed by cause and effect. Over a century later,

Jean Piaget reimagined a similar concept in psychology. He used the term *schemas* (the more common plural in modern psychology) to describe cognitive structures that children actively construct and adapt as they interact with the world. While Piaget's usage emerged independently of Kant's, both perspectives highlight the same essential insight: understanding depends on internal mental frameworks that bridge abstract concepts and lived experience. (Kant, 1781; Piaget, 1970)

Neuroscientific research nowadays indicates that schemata may have a physical basis in the brain's information storage patterns. As we learn, groups of neurons that frequently activate together begin to represent high-level concepts, creating what some neuroscientists call neural manifolds or low-dimensional representations within neural activity. (Langdon, Genkin and Engel, 2023) A neural manifold refers to the finding that neural firing patterns for related experiences often occupy a structured, smaller subspace of all possible activity patterns rather than being randomly distributed. Put more simply, a *manifold* is just a fancy word for an organized pattern in neural firing.

Think of neural manifolds as the brain creating simplified 'movies' of our experiences. Just as filmmakers create animations that capture complex movements with just the essential elements—preserving what matters while dropping unnecessary details—neural manifolds track essential patterns in brain activity as experiences unfold. This compression of neural information helps your brain efficiently categorize and connect similar experiences. For example, when you recall different restaurant visits, consistent neural patterns activate in your medial prefrontal cortex and hippocampus, regardless of which specific restaurant you're remembering. These streamlined neural patterns form the foundation of your "restaurant schema," a mental shorthand that helps you navigate dining experiences without processing every detail from scratch.

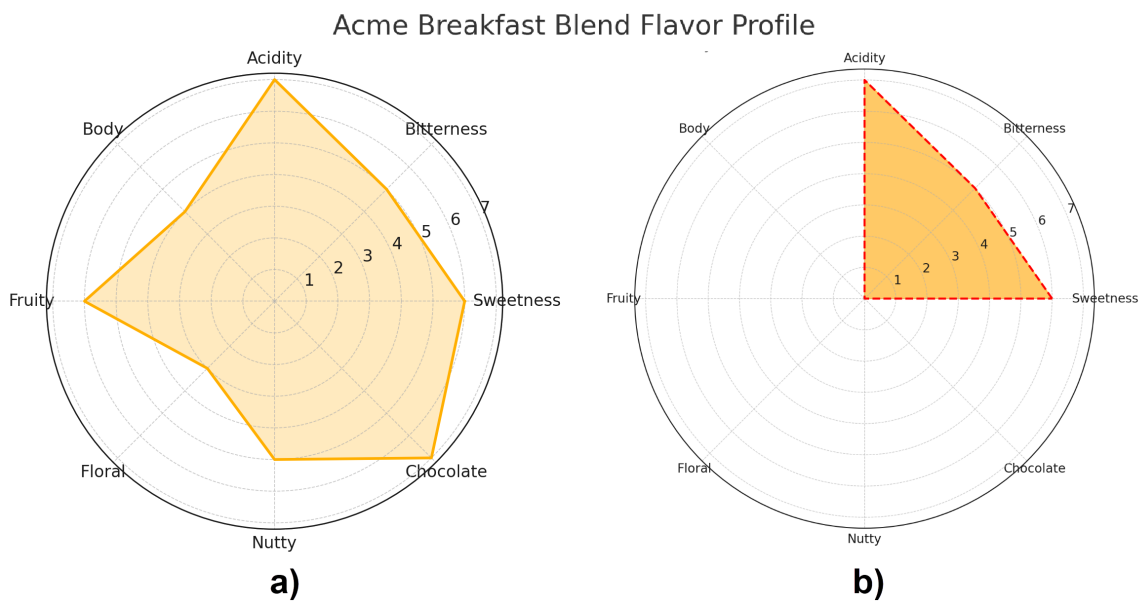

**Figure 6:** Figure 6: Dimensionality reduction. (a) The different tastes of coffee can be thought of as "dimensions"—this Acme blend has eight dimensions, astronomically simpler than neural information in the brain, which spans hundreds of trillions of synapses (more connections than stars in the Milky Way galaxy). (b) Dimensionality reduction simplifies complexity by focusing on only essential features—here, just acidity, bitterness, and sweetness. Our brain performs inconceivably greater compression, creating "neural

shadows" that distill this vast complexity into manageable representations, preserving essential patterns while filtering out enormous amounts of noise.

This dynamic transformation of memory—from richly detailed to more abstract and generalized—appears to sweep forward along the brain's architecture. Robin and Moscovitch describe this as a continuum of representation: initial memories are grounded in detailed perceptual features supported by posterior regions of the hippocampus and sensory cortices. As the memory becomes more generalized and gist-like, activity shifts toward the anterior hippocampus. With continued use, abstraction, and integration across experiences, these representations are increasingly supported by the ventromedial prefrontal cortex, where schematic knowledge is stored. Rather than replacing one another, these forms of memory coexist, allowing the brain to flexibly access different levels of representation depending on the situation. (Robin and Moscovitch, 2017) This back-to-front sweep—from raw perceptual input to generalized knowledge—parallels the progression described in Global Workspace Theory, in which specialized modules feed information forward to more central, integrative hubs that coordinate task-relevant cognition. (VanRullen and Kanai, 2021)

From a neural manifolds perspective, this transformation involves progressive dimensional reduction—detailed memories preserve specific perceptual features, gists distill essential elements of single episodes, and schemata form compressed representations that capture regularities across multiple experiences. Langdon and colleagues suggest these patterns aren't merely statistical abstractions but reflect the physical structure of neural circuits.

The transformation of scattered, high-dimensional neural activity into these efficient, lower-dimensional manifolds happens gradually through learning and practice. As we repeatedly engage with related concepts, our neural activity becomes more organized, converging on optimized pathways that require less cognitive effort. This efficiency explains why experts can process information in their domain more quickly than novices—their brains have developed well-defined neural circuits for familiar concepts.

Interleaved practice – mixing different yet related topics or skills in a learning session instead of tackling one at a time – has been shown to strengthen long-term retention and yield more flexible, transferable knowledge. By forcing the brain to continually retrieve and differentiate between multiple concepts, this approach reinforces the distinct neural manifolds underlying each skill. Intriguingly, artificial neural networks trained on interleaved tasks show a similar advantage: they consolidate skills more robustly and forget less. This parallel underscores that both the human brain and artificial neural networks benefit from interleaved training. (Buzsaki, McKenzie and Davachi, 2022; Foster *et al.*, 2019; Carvalho and Goldstone, 2019; Mayo *et al.*, 2023)

Even in brain regions where individual neurons behave in complex and highly variable ways, the combined activity of many neurons can form organized patterns that create neural manifolds. According to Langdon and colleagues, these low-dimensional manifolds arise from simple connectivity principles embedded within the complex neural architecture. It's like how a symphony orchestra creates coherent music from many individual instruments—the neural manifold is the recognizable "melody" that emerges when many neurons follow underlying connectivity rules. Scientists have directly observed this relationship between circuit connectivity and manifold structure in brain systems for navigation, and similar principles likely apply to how memory manifolds form and transform. (Langdon, Genkin and Engel, 2023)

**5.3 Types and Applications of Schemata**

David Rumelhart, a foundational figure in cognitive science and early neural network modeling, helped formalize the concept of a "schema"—a flexible mental framework for interpreting experience, guiding memory, and shaping understanding. His model describes knowledge as hierarchically organized, with each schema composed of embedded subschemata that represent progressively more specific components of a concept (Rumelhart, 1980). While Robin and Moscovitch describe how individual memories transform over time—from richly detailed to schematic as they progress across brain regions—Rumelhart's account emphasizes how schemata themselves evolve through experience. In his framework, knowledge structures adapt through accretion, tuning, and restructuring: they can become more refined, generalized, or reorganized as new information is encountered. Thus, both perspectives describe transformation, but at different levels—Robin and Moscovitch focus on changes within a memory trace; Rumelhart focuses on how the cognitive frameworks we use to interpret experiences are themselves reshaped over time. Together, these views reinforce the idea that developing robust internal schemata is not just helpful for learning—it is structurally and functionally essential for efficient thinking, creative insight, and expert performance.

If schemata help organize our experiences and guide our thinking, how do they actually work in the brain—and how do they change with use? While schemata are abstract mental frameworks, they correspond to patterns of neural activity that can grow stronger, more efficient, or more interconnected over time. These neural manifolds aren't static—they evolve as we learn. Each new experience that fits an existing pattern strengthens it, while mismatches can lead to restructuring or expansion of the manifold. This kind of adaptive refinement echoes Piaget's classic concepts of assimilation and accommodation—processes by which new information is either integrated into existing mental structures or prompts those structures to change.

Hebbian learning—where neurons that repeatedly activate together form stronger connections—helps explain how this strengthening occurs at the neural level, though schemata themselves remain abstract structures. By continually refining these neural patterns, the brain's prefrontal cortex acts like an efficient data compressor, creating simplified "maps" of knowledge that capture key relationships while discarding irrelevant details. This process directly supports cognitive category formation and classification, allowing us to recognize meaningful patterns across similar experiences and concepts quickly and accurately. (Bein and Niv, 2025; Hebscher *et al.*, 2019)

Recent research suggests that this kind of neural efficiency may begin even earlier than expected. During active maintenance in working memory, the brain appears to reshape incoming information into simplified, low-dimensional formats that anticipate future decisions. These transient representations may reflect existing schemata—and, in turn, help refine them—gradually tuning the brain's architecture for faster, more flexible thinking. (Wojcik et al, 2025).

Different types of schemata serve various cognitive functions. Spatial schemata help us navigate physical environments by encoding common patterns in spatial layouts. As psychology researcher Delaram Farzanfar (2022) and her colleagues have observed, the brain forms "spatial schemata" that capture common patterns across similar environments. Just as we have a restaurant schema, we develop schemata for typical spatial layouts like modern cities or shopping malls. These spatial schemata exist in a hierarchy: detailed "cognitive maps" represent specific places (like your hometown), "spatial gists" capture essential elements while dropping details, and overarching "spatial schemata" represent common patterns that apply across many environments. This organization enables efficient navigation in new places that share features with previously experienced environments. When we rely too heavily on GPS and navigation apps, we may prevent the formation of these robust spatial schemata, leaving us disoriented when technology fails.

Schema theory also suggests three main reasons why comprehension or recall might fail: (1) a person may lack the appropriate schemata entirely; (2) the necessary schemata may exist but insufficient cues are present to activate them; or (3) the person may activate a consistent but incorrect schema, leading to misunderstanding. These failure modes help explain why knowledge that exists 'out there' rather than being internalized often proves inadequate when confronting novel problems.

We also develop procedural schemata that guide sequences of actions. A musician develops schemata for chord progressions, fingering techniques, and typical melodic patterns. A mathematician builds schemata for common problem-solving approaches and proof strategies. These procedural schemata connect with the declarative-to-procedural transition described in Section 3, where frequently-used knowledge gradually transforms into skill-like expertise stored in procedural memory systems.

In the absence of schemata—when new information doesn't connect to existing knowledge—the brain struggles to process or make sense of it. Without schemata to simplify and structure the information, the brain must handle each new piece in its full complexity, often failing to find any meaningful pattern at all. For example, someone who learns programming by merely copying code from the internet might solve immediate problems but won't develop a deeper mental model. When encountering novel challenges, they have no framework to guide them—just scattered, disconnected fragments of information.

What's particularly important is how these schemata integrate with prediction processing in the brain. Our brains constantly generate predictions about the world, and when those predictions fail, we experience prediction errors. These prediction errors drive us to update our mental models—or in neural terms, refine our manifolds. This updating process is especially powerful when it leads to a sudden insight or "aha" moment, creating a strong memory advantage that enhances learning. (Becker and Cabeza, 2024)

The brain's ability to organize information into these efficient structures allows us to navigate complex environments without having to process every detail anew. These neural organization principles represent the brain's remarkable capacity to identify patterns, make predictions, and learn efficiently from experience.

**5.4 Schema Formation and Learning Optimization**

Often, creative breakthroughs occur precisely after we feel most stuck or confused. At these moments, learners are primed to reorganize their understanding, producing the well-known 'Aha!' experience." Indeed, cognitive scientists refer to challenges that feel difficult in the moment but facilitate deeper, lasting understanding as "desirable difficulties."(Bjork and Bjork, 2011) Unlike deliberate practice, which systematically targets specific skills through structured feedback, desirable difficulties leverage cognitive struggle to deepen comprehension and enhance retention. (Ericsson, 2009)

Recent neuroscience has identified an optimal challenge level for effective learning. The "Eighty Five Percent Rule" shows that learning is best when students achieve about 85% accuracy during practice. (Wilson *et al.*, 2019) Echoing Vygotsky's classic "zone of proximal development," this sweet spot—neither too easy nor too difficult—helps neural networks efficiently form and strengthen connections. Educators can use this insight as a practical guide, balancing task difficulty to encourage strong neural manifold development without causing cognitive overload.

High-dimensional neural spaces present both opportunity and challenge for learners. While they enable sophisticated conceptual integration, they also create vast landscapes of potential solutions that can overwhelm students, particularly those with mathematical learning difficulties like dyscalculia. (Research suggests that rather than too few, these students often maintain too *many* possible cognitive pathways, lacking the structured organization needed to efficiently navigate toward correct solutions. (Geary, Berch and Koepke, 2019, p. 17))

This insight challenges popular discovery-based learning approaches. Expecting a child to discover mathematical principles independently is like placing them in a car and saying, "Figure out how to drive." The myriad possible actions and interdependent operations make it virtually impossible for novices to navigate to effective solutions without structured guidance. Just as driving instructors constrain the learning space by focusing attention on relevant controls, mathematical instruction provides scaffolding that helps students navigate complex conceptual spaces efficiently.

The core issue isn't a lack of neural connections but insufficient pruning and organization of those connections into optimized pathways. Students with dyscalculia often engage multiple competing neural pathways simultaneously, creating cognitive interference that impedes problem-solving. Guided instruction serves as a critical scaffold—constraining the high-dimensional space, directing learners toward productive pathways, and reducing cognitive load from irrelevant dimensions.

This explains why external aids like calculators, without corresponding internal schema development, fail to produce lasting mathematical proficiency. Calculators provide correct answers but don't develop the optimized neural manifolds necessary for intuitive understanding. Through appropriate practice maintaining the 85% success rate, students develop structured neural manifolds that allow efficient navigation of mathematical concepts, reducing the dimensionality of the problem space while preserving essential relationships.

When students practice multiplication facts, they aren't merely memorizing arbitrary associations but creating optimized neural patterns that represent numerical relationships. These structured manifolds become the foundation for recognizing mathematical patterns—a far more powerful cognitive tool than an overwhelming space of undifferentiated possibilities.

## 5.5 Cognitive Offloading and Schema Disruption

Building a schema requires repeatedly encountering related information and actively connecting new details into existing knowledge. If each new piece of information is simply looked up without integrating it into memory, we fail to build lasting schemata. This is like trying to complete a jigsaw puzzle by checking the picture on the box for every single piece, rather than developing a sense of how the pieces fit together. You might correctly place individual pieces, but you won't internalize the overall picture. Over time, relying on external lookups for every problem makes it hard to recognize common patterns—each problem seems entirely new because you haven't built the underlying schema. Constantly looking things up becomes mentally exhausting and inefficient, compared to quickly recalling what you've already learned.

From a neural perspective, without sustained internal engagement, the brain might not solidify a low-dimensional manifold for the task. Each instance might be processed in a more onerous, high-dimensional way (basically, treating it as new) because the shortcut of a schema isn't available. This connects to the fundamental nature of schemata as active computational processes that evaluate the quality of their fit to available data. Without well-developed internal schemata, the brain loses its ability to efficiently determine whether and to what degree it can account for a given situation, forcing it to repeatedly process familiar information as if it were new. In the long run, this can make cognition less efficient and flexible. For example, learning a programming language by copying code from the internet for every task (without understanding) might get immediate results, but the person never develops a schema of how the code works. When faced with a slightly novel problem, they're stuck—there's no mental model to adapt, just a bag of fragments. By contrast, someone who internalized coding concepts can handle new problems by drawing analogies to things they already understand, thanks to robust schemata.

Research underscores these concerns. One review of technology-based cognitive offloading observes that offloading information to digital storage often results in only the gist of the information (or its location) being retained in the learner's mind. In essence, people remember *that* something can be looked up and roughly what it's about, but not the details. These retained "biological pointers" to external information create an *illusion* of knowledge—we feel like we know it because we know where to get it—but our actual cognitive schema remains impoverished. (Skulmowski, 2023)

Moreover, if learners attribute low value to deeply learning content (because it's readily accessible outside), they invest less effort, creating a vicious cycle where shallow engagement leads to shallow encoding. Offloading may also interfere with the normal formation of integrated memory networks by constantly disrupting the flow of information into working memory and long-term memory. High extraneous cognitive load (e.g. dealing with a distracting interface or juggling lookups) can prevent information from even entering working memory, much less being encoded in long-term memory. In such cases, *only* a fragmentary trace (a pointer or feeling of "I can always find this later") is stored. From the standpoint of schema theory, this is a poor substitute; a pointer is not a schema. A schema enriches understanding by linking concepts, whereas a pointer simply says "info exists somewhere." A library is useless to someone who doesn't know enough to find or contextualize what's in the books.

## 5.6 Neural Evidence and Implications for Learning

Neural evidence also hints at how incomplete internalization might affect problem-solving. Studies of insight and memory (like those by Becker et al.) show that the medial prefrontal cortex (mPFC)—a brain region crucial for integrating information and supporting higher cognitive functions—rapidly integrates new information with existing schemata in neocortex when a person learns with insight. If one has no schema (because prior knowledge is absent or unorganized), that integration process likely fails or is less efficient. The person might learn the isolated fact but won't connect it to broader understanding. Similarly, bridging the gap between neural manifolds and neural circuitry is crucial for understanding how learned knowledge is represented. This relationship between abstract manifolds and physical brain networks explains why well-organized knowledge is more efficiently processed. When a learner lacks structured manifolds, their neural activity remains disorganized and inefficient—the brain processes each task as if encountering it for the first time rather than leveraging established patterns. In essence, the brain stays in "novice mode," using energy-intensive, widespread neural activity for tasks that an expert's brain would handle with a streamlined, specialized network. (Becker and Cabeza, 2025; Langdon, Genkin and Engel, 2023)

In practical terms, one way to think of schemata and neural manifolds is that they are the internal mental maps we use to navigate domains of knowledge. Relying excessively on external GPS (so to speak) means we never draw our own map. When the GPS is gone or gives incomplete info, we're lost. Conversely, if we have a solid mental map, external tools can still aid us (just as a driver with local knowledge can still use a GPS for efficiency), but we are not helpless without them because we've built the internal schema that lets us think independently.

Recent neuroscience provides compelling evidence for this concept. A 2025 study of flight trainees demonstrated how specialized training creates measurable changes in brain network connectivity and efficiency. (Ye, Ba and Yan, 2025) Researchers found that after completing intensive flight training, trainees exhibited significantly enhanced functional connectivity between the Central Executive Network and Default Mode Network—essential systems for cognitive control and information integration. Moreover, these trainees spent more time in brain states characterized by this efficient connectivity, and these temporal metrics directly correlated with performance on cognitive tests measuring flexibility and visual processing. This suggests that intensive practice not only strengthens connections between brain networks but also creates more stable, efficient neural states—essentially the neural manifestation of well-formed schemata that allow experts to process information more efficiently than novices.

> **Sidebar: Shadows of Thought: How Generative AI Mirrors the Brain**
>
> Generative AI models like ChatGPT share striking parallels with how human brains form schemata. Both simplify vast amounts of information into manageable patterns—brains through neural manifolds, AI through structured representations learned from extensive data. These simplified structures act like "shadows," capturing essential features by reducing complex, high-dimensional data into simpler forms. This helps explain how both brains and generative AI efficiently recognize patterns, predict outcomes, and solve problems.
>
> Recent research takes this parallel even further. Gerald Pao's *generative manifold networks*, for example, use manifold-like structures that enable AI to predict neural responses and generate realistic behaviors not explicitly programmed (Pao et al., 2021). Markus Buehler's research (2024) goes even deeper, demonstrating that sophisticated AI systems actively and autonomously explore knowledge by iteratively questioning, reflecting, and reasoning *within their own evolving knowledge networks*—continuously discovering novel connections and insights. This proactive internal exploration mirrors how human learners actively build robust schemata through repeated reflection and reasoning within their own neural architecture.
>
> This insight underscores a central theme of this chapter: the true value of knowledge lies not merely in external accessibility, but in the active internal integration of that knowledge within our cognitive architecture. Excessive reliance on external tools for memory bypasses this critical internal process, weakening precisely those mental structures—schemata and neural manifolds—that foster deep understanding and creative insight in both humans and advanced artificial systems. (Sejnowski, 2024)

The implications for education are significant. When we understand how schemata form and function neurologically, we can design learning experiences that facilitate their development rather than hinder it. Effective education should balance the use of external tools with opportunities for students to internalize key knowledge and develop rich, interconnected schemata. This balance ensures that technology enhances learning rather than creating dependence and cognitive weakness.

# 6. Cognitive Offloading and Learning Trends

### 6.1 Historical Shift: From Memorization to "Look It Up"

Mid-20th century classrooms emphasized drills, repetition, and frequent testing of factual knowledge—techniques derided by some as "drill and kill." By the late 20th century, education pivoted toward critical thinking skills and away from memorization, adopting the mantra "you can always look it up." As Daisy Christodoulou, author of *Seven Myths about Education*, notes:

> "The biggest contemporary myth about education is the idea that knowledge no longer matters. People now say that know-how is more important than knowledge, since children don't need to know things they can look up on their smartphones at any time."(Christodoulou, 2015)

This shift coincided with the advent of electronic calculators in the 1970s and personal computers and the internet in the 1980s and 1990s, enabling unprecedented cognitive offloading—the use of physical action to alter information processing requirements to reduce cognitive demand. Educators increasingly argued against memorization of content that could be readily accessed or calculated by machines, positioning this approach as freeing up time for "higher-order skills." This recommendation often found a receptive audience since, as research shows, thinking itself can be unpleasant for many people. (David, Vassena and Bijleveld, 2024)

This shift mirrors a recurring pattern in schooling, where practices that appear more intellectual and abstract (like teaching "critical thinking skills") gain status over seemingly basic practices (like memorization), regardless of their actual cognitive foundation or effectiveness. (Labaree, 2006)This shift away from memorization was fueled by the persistent status struggles of colleges of education within the academic hierarchy. As Labaree (2006) details, facing perceptions of lower rigor and lacking primary authority over subject matters, education schools often sought legitimacy by embracing sophisticated pedagogical theories. Progressive ideals, emphasizing student-centered learning processes and higher-order thinking over direct knowledge transmission, offered a seemingly more 'academic' mission than focusing on foundational content mastery and memorization, potentially aligning better with university norms while distancing the field from its lower-status training roots.

### 6.2 Quantifying the Shift: Language Analysis

To track this pedagogical evolution, we (the authors—not the omniscient 'we' that, yes, has snuck into this very chapter) analyzed phrases associated with less emphasis on memorization in published literature over time (Figure 7).

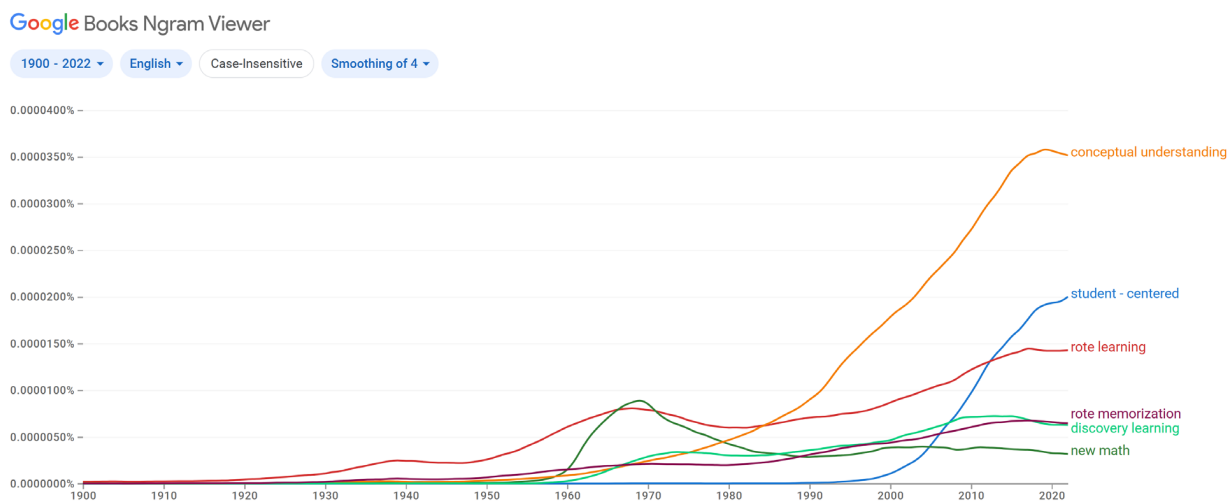

**Figure 7.** Trends in education-related phrases frequently associated with diminished emphasis on memorization. The colored lines (affiliated with the y-axis) indicate the frequency of each phrase in published books; the x-axis shows years from 1900 to 2020. Note that the phrase "conceptual understanding" is included because many educators have historically treated it as distinct from or even opposed to memorization, despite evidence showing memorization can support conceptual understanding.

Our sentiment analysis of the word "rote" in American books and publications shows how attitudes changed over time.[1] In the 19th and early 20th centuries, the term was largely neutral or slightly positive, appearing in educational contexts without negative connotations. However, by the 1930s–1950s, "rote" began to carry a more rigid implication, reflecting early critiques of mechanical memorization in learning. From the 1960s onward, sentiment toward "rote" declined sharply, coinciding with shifts in educational philosophy that prioritized critical thinking over rote memorization. This linguistic shift paralleled the rise of technologies that could store information externally (see Figure 8).

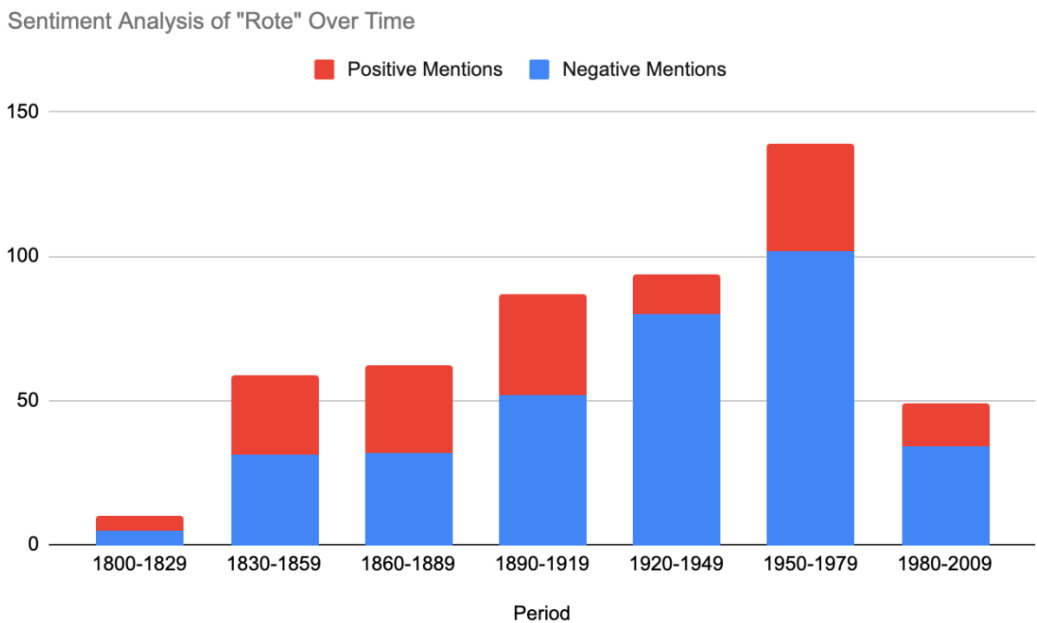

**Figure 8:** A sentiment analysis of the term "rote" in the Corpus of Historical American English (COHA) revealed a distinct evolution in the perception of "rote" across different periods.

### 6.3 The Cognitive Consequences of Offloading

However, as we saw in previous sections on schemata and procedural memory, completely eliminating foundational memorization can undermine those very higher-order skills. Research shows that offloading information doesn't just externalize storage – it transforms our cognitive

---

[1] Our analysis was conducted in the Corpus of Historical American English (COHA) using a hybrid approach, combining rule-based keyword filtering with Python-powered sentiment analysis. This allowed us to track shifts in connotation over time with both quantitative insights and contextual relevance. Given that this study focused on a single keyword, we implemented Python scripting within KNIME to automate sentiment classification, leveraging transformer-based NLP models to assess sentiment trends across historical texts.

processes. Risko and Gilbert (2016) propose a three-component framework for this cycle: (1) our decision to use external tools is influenced by our metacognitive evaluation of our own abilities, (2) the act of offloading then affects those metacognitive judgments, and (3) offloading directly impacts our cognitive capabilities. This creates a self-reinforcing cycle where offloading can lead to increased dependence on external tools. (Risko and Gilbert, 2016)

When students repeatedly outsource cognitive processes to technology, they develop "biological pointers"—remembering where to find information rather than the information itself. While seemingly efficient, this creates an illusion of knowledge that undermines schema formation and deeper understanding.

### 6.4 The Knowledge Paradox: Hirsch's Warning

This concern isn't new. E.D. Hirsch Jr. warned that without domain knowledge, critical thinking falters. In his 2000 article "'You Can Always Look It Up'... Or Can You?"(Hirsch, 2000), he argued that broad factual knowledge enables effective skill use – contradicting the notion that knowledge and skills are separate. Without background knowledge, students can't evaluate sources or recognize whether information is plausible. In Hirsch's words, "those who repudiate a fact-filled curriculum on the grounds that kids can always look things up miss the paradox that de-emphasizing factual knowledge actually hinders children from learning." A novice doesn't know what they don't know – without stored knowledge, they can't formulate proper questions or search strategies.

Empirical studies support these arguments. In one classic experiment(Miller and Gildea, 1987), children told to look up unfamiliar words produced nonsensical sentences. Why? They misinterpreted definitions without vocabulary context, and the lookup process itself created cognitive overload. The children would have learned more effectively if taught the words in context rather than being told to "look it up." This illustrates a crucial principle: external resources benefit those who already possess internal knowledge. As Hirsch put it, "to be able to use that information [from the Internet] – to absorb it, to add to our knowledge – we must already possess a storehouse of knowledge."

### 6.5 Digital Amnesia: The Google Effect

Another line of research has examined what is sometimes called the Google effect or "digital amnesia." Psychologist Betsy Sparrow and colleagues found that when we expect future access to information, we remember where to find it rather than the information itself. (Sparrow, Liu and Wegner, 2011) This isn't entirely negative – it's a form of adaptive memory management by the brain – but it underscores that offloading can change what we remember.

In moderation, this division of labor between internal and external memory is efficient. But applied indiscriminately, it can overload working memory with biological pointers rather than direct knowledge use. Imagine trying to write an essay while looking up every fact or word – the flow of thought would be constantly disrupted. Working memory, which is limited, gets consumed by the mechanics of search instead of the synthesis of ideas. Ironically, excessive cognitive offloading can increase cognitive load during complex tasks. Tools do lighten the load for isolated tasks (calculating a sum, finding a date), but if reliance becomes too high, the tool use itself becomes an overhead in any integrated task.

# 7. The Flynn Effect, Its Reversal, and Possible Links to Memory Practices

The cognitive offloading trends described in the previous section raise an important question: Could these widespread shifts in how we store and access information have measurable effects on cognitive abilities at a population level? If educational practices and daily habits increasingly favor external memory storage over internal knowledge building, we might expect to see these changes reflected in standardized measures of cognitive performance. The Flynn Effect and its recent reversal offer a revealing window into this possibility—a natural experiment in how changing approaches to knowledge might influence measurable intelligence.

## 7.1 The Rise, Fall, and Potential Causes of IQ Scores

Population-wide IQ scores provide insights into how memory practices affect cognitive abilities. Throughout most of the 20th century, IQ scores steadily climbed—a phenomenon called the Flynn Effect. Two major studies showed scores rising about three points per decade. (Pietschnig and Voracek, 2015; Trahan *et al.*, 2014) People worldwide were getting "smarter" on these tests, thanks to better schools, nutrition, health, and more complex environments.

Then something unexpected happened. Starting with people born after the mid-1970s, researchers spotted a reversal—particularly in wealthy countries. Norwegian data showed scores dropping by up to seven points per generation. (Bratsberg and Rogeberg, 2018) Denmark, Britain, France, the Netherlands, and others showed similar declines. (Teasdale and Owen, 2008; Dutton, van der Linden and Lynn, 2016)

Interestingly, IQ scores continue rising in developing countries as education and living standards improve. The decline appears mainly in advanced nations. What changed? Most revealing is that IQ declines appear within families—younger siblings scoring lower on average than their older siblings, effectively ruling out simple genetic or demographic explanations. Researchers are investigating whether shifts in education, technology use, and cognitive habits might explain why the decades-long upward trend has reversed in wealthier societies.

## 7.2 Timing and Potential Causes

Researchers have explored numerous hypotheses for the Flynn Effect's reversal—from declining test motivation to environmental pollutants to nutritional shifts. The evidence points primarily to environmental and cultural factors rather than genetics. (Teasdale and Owen, 2008)

The timing of this reversal offers a compelling clue. The cohorts born after approximately 1975 were educated during the 1980s–2000s—precisely when Western education shifted away from memorization toward "learning to learn"—an admirable goal pursued by educators who, ironically, had limited understanding of how learning actually works in the brain. This period also saw digital technology—calculators, computers, and eventually the internet—become fixtures of daily life. A new mindset emerged: "Why memorize knowledge or methods when you can always look them up or use a device?"

As schools emphasized abstract thinking skills over content knowledge, information and computation became instantly accessible at the push of a button. This raises a critical question: Were the cognitive skills that boost IQ performance—mental arithmetic, broad vocabulary, general knowledge—exercised less in this environment? The chronological correlation between rising digital dependence and plateauing IQ scores is striking, though correlation isn't causation. Multiple factors likely interweave, making it difficult to isolate a single cause.

Could the continued rise of the Flynn Effect in developing countries, contrasted with its decline in wealthier nations, relate to Western acceptance of constructivist approaches that de-emphasize memorization? Evidence suggests that when wealthier, high-performing countries, as for example, Taiwan, experiment with constructivism, their math scores decline. As they might then move away from constructivist and discovery learning approaches, mathematics performance quickly improves. (Lai, 2024; Lin *et al.*, 2014; Oakley, Chen and Johnston, 2025) This dynamic exemplifies Pedagogical Attribution Displacement (PAD), where educators may falsely attribute student success to fashion pedagogical approaches, such as constructivism, while underappreciating the foundational knowledge established through prior traditional instruction. Such misattributions arguably stem from the very challenge Bjork, Dunlosky & Kornell (2013, p.417) identified: "*people often have a faulty mental model of how they learn and remember, making them prone to both misassessing and mismanaging their own learning*." When educators operate with unjustified teaching beliefs(Double *et al.*, 2020)—perhaps undervaluing the deep structure provided by earlier instruction—they risk misinterpreting the trustworthy source of student achievement, potentially contributing to complex narratives surrounding educational reforms and international rankings (e.g., in Finland, see Sahlgren, 2015).

There is a clear bidirectional link between mathematics and IQ. It's not just that smarter kids can learn math better. Learning math, it seems actually helps make you smarter—at least from an IQ test perspective. (Song and Su, 2022) Less effective methods for teaching mathematics, in other words, could lead to the decline of IQ in entire nations.

Recent reporting on cognitive performance declines, such as that by Burn-Murdoch (2025) in the *Financial Times*, often attributes these trends primarily to technology distractions and shifting media consumption patterns. While these environmental factors may certainly play a role, they may obscure a more fundamental issue: educational approaches that have systematically devalued the memory foundation necessary for higher-order thinking. Educational institutions readily embrace such technology-focused explanations, as they conveniently shift attention away from examining how pedagogical trends might be contributing to these declining outcomes. After all, it's far easier to blame smartphones than to reconsider cherished educational reforms that have been promoted as progressive advancements.

**7.3 Educational Shifts, Cognitive Exercise, and Technology Balance**

How do educational practices influence specific cognitive abilities measured by IQ tests? This question is central to understanding the Flynn Effect reversal.

IQ assessments measure various cognitive abilities: arithmetic, vocabulary, general knowledge, memory span, and abstract problem-solving. While designed to gauge general intelligence, these tests inevitably reflect practiced skills. Someone who has memorized math facts solves arithmetic problems faster; someone who reads widely knows more vocabulary. Test performance reflects which cognitive skills have been exercised.

Studies consistently show knowledge-based and verbal abilities showing the greatest declines, while processing speed and some non-verbal abilities remain stable or improved. In France, Vocabulary, Comprehension, and Information subtests showed the largest drops, while processing speed increased slightly and memory span remained unchanged (Dutton and Lynn, 2015). This suggests recent cohorts know fewer words and facts than their late-1990s counterparts.

American adults show significant vocabulary declines between the 1970s and 2010s (0.5 to 1.3 IQ points per decade), regardless of education level (Dworak, Revelle and Condon, 2023). These findings suggest reduced reading and increased internet reliance have affected vocabulary and general knowledge. By contrast, tasks less dependent on acquired knowledge haven't declined equivalently.

This pattern contrasts with the original Flynn Effect, where mid-20th-century IQ gains were strongest in non-verbal reasoning, while verbal and knowledge-based skills rose modestly (Dutton, van der Linden and Lynn, 2016). Today, knowledge-centric skills show relative decline precisely when modern education and habits have deemphasized these areas.

James Flynn noted that while young children's environments became more cognitively enriched (with educational shows and toys), teenage environments might have become less intellectually stimulating, with curricula focusing less on content and teens spending more time on screens and entertainment rather than reading and studying.

These findings align with our understanding of memory systems: frequently offloading cognitive work to devices may cause certain "mental muscles" to atrophy. Students expected to internalize core knowledge develop sharper recall and better mental organization, while those constantly relying on external aids may never develop robust internal knowledge structures.

The Flynn Effect's rise and fall have complex origins that resist simplistic explanations. Nevertheless, multiple analyses confirm recent IQ declines reflect real drops in specific cognitive capabilities. If people in 1960 memorized more and performed more unaided mental problem-solving than people in 2020, certain mental proficiencies would naturally differ between these populations, suggesting that how we train memory matters at a societal level.

A compelling interpretation connects these trends: Mid-20th-century education emphasized content knowledge and coincided with unprecedented IQ gains. Late-20th-century education claimed to prioritize "learning how to learn" over memorization but implemented methods that neuroscience now reveals were actively hampering effective learning—coinciding with stagnating or declining IQ scores. This correlation demands reconsideration of approaches that devalued memory under the misguided assumption that it wasn't essential for deeper thinking. (Novak and Gowin, 1984)

Evidence shows heavy reliance on digital aids correlates with diminished analytical reasoning, highlighting the importance of maintaining internal cognitive habits alongside technological conveniences. If we want to preserve cognitive gains, we shouldn't abandon training the "memory muscle." How we teach young minds has consequences that reverberate through our collective cognitive capabilities. (Barr *et al.*, 2015)

**Educational Giants of the Past: What they got right—and wrong**

Many great educational leaders of the past were intellectually brilliant; paradoxically, this brilliance may have sometimes blinded them to the reality that most students benefit greatly from explicit, structured instruction, particularly for abstract, biologically secondary knowledge. Indeed, all these pioneers left double-edged legacies, advancing educational thinking in crucial ways while simultaneously creating significant roadblocks that delayed progress. Below, we briefly explore these educators' key insights and missteps in light of modern neuroscience and cognitive load theory.

- **Benjamin Bloom (1913-1999)** transformed education through his Taxonomy of Educational Objectives and Mastery Learning approach. His renowned "two-sigma problem" revealed that one-on-one tutoring paired with mastery techniques can dramatically boost learning—an insight now echoed by neuroscience, which shows that targeted feedback and deliberate practice strengthen neural networks.
    Yet modern neuroscience exposes key limitations in how Bloom's frameworks have often been interpreted. His taxonomy implies a rigid hierarchy of cognitive skills, whereas the brain's operations are far more integrated. Higher-order thinking—such as analysis and creativity—relies not on separable faculties but on richly interconnected knowledge networks. Similarly, Bloom's taxonomy divided learning into cognitive, affective, and psychomotor domains—an approach he acknowledged was pedagogically useful but somewhat artificial. He recognized that these aspects of learning overlap in practice, yet maintained their separation to help educators plan and assess instruction. Neuroscience now confirms that such separations are not just artificial but misleading: emotion and cognition continuously interact, jointly shaping memory, reasoning, and attention.
    While Bloom's tutoring research underscored the power of personalized learning, later studies have rarely reproduced the dramatic two-sigma gains. The two-sigma result is not a statistical fluke, but it requires a constellation of best practices rarely achieved together, even one-on-one. It's individually possible, but systemically elusive. Neuroscience shows that lasting learning depends on active retrieval, timely feedback, adaptive pacing, and emotional investment—elements central to Bloom's original findings, yet too often missing in educational reform.
- **Jerome Bruner (1915–2016)**, an influential psychologist known for advocating discovery learning, argued that students should independently explore and discover concepts. While beneficial when carefully structured, neuroscience reveals Bruner's fully unguided discovery methods can overwhelm learners' limited working memory, particularly when dealing with biologically secondary material. Bruner thus underestimated the vital role structured guidance plays in building robust, biologically secondary knowledge schemata.
- **John Dewey (1859–1952)**, an American philosopher and educational reformer, pioneered experiential education, emphasizing reflection as central to deep learning. Dewey largely dismissed the value of rote memorization, assuming that experience combined with reflection alone would naturally consolidate knowledge. Ironically, modern neuroscience confirms that explicit practice in recalling information (yes, even good old-fashioned memorization) is crucial—no matter how unfashionable—for most learners to reliably encode biologically secondary information into long-term memory.

- **Maria Montessori (1870–1952)**, Italian physician and educational innovator who developed a child-centered approach emphasizing self-directed, hands-on learning. Her methods proved extraordinarily effective for young children in mastering foundational (biologically primary) skills. Montessori believed this student-driven, concrete learning approach could likewise be applied to more advanced knowledge. In doing so, however, she underestimated the need for explicit instruction as learners grew older and content became more abstract. Modern research shows that some guided teaching and deliberate practice are indispensable for efficiently encoding complex (biologically secondary) academic concepts as children mature.
- **Jean Piaget (1896–1980)**, a pioneering Swiss psychologist, correctly theorized cognitive development arising from "cognitive disequilibrium"—the mismatch between expectation and reality. Piaget rightly understood that learners naturally reorganize mental models regarding biologically primary domains (such as social interactions or native language skills). However, he mistakenly extended this assumption to biologically secondary knowledge, presuming learners could spontaneously reorganize their mental models in these areas without explicit guidance. Current neuroscience emphasizes that structured instruction and deliberate feedback are vital for reliably developing stable and accurate schemata, particularly for abstract biologically secondary knowledge.
- **B.F. Skinner (1904–1990)**, an influential American behaviorist, insisted that learning occurs through operant conditioning—systematic reinforcement and immediate feedback shaping behavior. Skinner's insights were so powerful that they inspired modern artificial intelligence methods like reinforcement learning, where algorithms learn via reward signals. However, Skinner treated the mind's inner workings as a "black box," strictly focusing on observable behaviors and outcomes. He dismissed internal cognitive processes as unobservable and irrelevant—precisely the processes, such as engrams, schemas, and manifolds, that neuroscience now recognizes as central to understanding and enhancing learning. Thus, Skinner's legacy, like so many of the greats in education, is paradoxical: he was correct about the critical role of reinforcement and feedback yet wrong in blocking investigations into the brain's internal mechanisms that are essential if we're to truly advance our understanding of how to learn and teach effectively.
- **Lev Vygotsky (1896–1934)**, a Russian psychologist who founded social constructivism, highlighted learning as inherently social, situated within a "Zone of Proximal Development." Vygotsky, though insightful, overestimated the sufficiency of social interaction alone for teaching biologically secondary knowledge. Neuroscience has clarified that explicit, structured instruction is crucial to seat abstract, biologically secondary concepts within long-term memory effectively.

# 8. Conclusion: Balancing Internal Memory and Digital Tools for Cognitive Health

## 8.1 The AI Challenge: Metacognitive Laziness

Humans naturally use external tools—like language, symbols, diagrams, and notes—to help think and communicate clearly. These external tools enhance our cognitive abilities by offloading memory and simplifying complex ideas. But as we've seen, relying too heavily on external aids can weaken our internal mental frameworks. Effective learning requires a balance: using external tools to support—not replace—the deep internal knowledge essential for genuine understanding. (Fernando *et al.*, 2024)

We began with a simple premise: human memory matters, even (or especially) in the digital era. Through this chapter, we explored how and why. We saw that the brain's learning mechanisms—from the dopamine-driven thrill of prediction errors during insight to the basal ganglia's slow burn of procedural memory formation—are designed to function with information that the brain holds and manipulates internally. When used properly, technology can *augment* these processes (for instance, by providing rich feedback or additional examples), but if used as a crutch to avoid mental effort, it can diminish cognitive outcomes and foster metacognitive laziness. Over-reliance on digital externalization carries several risks. It can rob us of the memorable impact of discovery, stunt the formation of fluent skills and intuition, disrupt the building of neural manifolds that make us efficient thinkers, and even contribute to broader cognitive stagnation or decline. In essence, an offloaded mind may become an under-exercised mind—one that increasingly lacks awareness of its own knowledge gaps.

The arrival of AI assistants like ChatGPT has intensified these concerns about cognitive offloading in education. Do AI systems that instantly provide answers or writing help enhance learning, or merely substitute for it? Recent research provides compelling evidence for the latter.

A revealing study titled "Beware of Metacognitive Laziness: Effects of Generative AI on Learning" compared college students writing essays with different types of assistance: ChatGPT, human tutoring, basic tools, or no help. (Fan *et al.*, 2024) The results were striking—ChatGPT-assisted students produced higher-quality essays but showed no knowledge improvement when tested later, and in some measures actually performed worse. These students displayed what researchers termed "metacognitive laziness"—fewer self-correcting behaviors and significantly less time reflecting on the material. They took the path of least resistance, letting the AI generate content without deeply engaging with the subject—effectively bypassing the prediction error mechanisms that create strong memory engrams.

Recent neurophysiological research from MIT supports this concern: learners who began writing with ChatGPT showed weaker connectivity in frontoparietal brain regions linked to focus and memory—and had more trouble recalling their own work. In contrast, those who wrote first and used AI only later retained more and showed stronger neural engagement. The timing of AI use, in short, may shape not just how we think, but whether we remember. (Kosmyna et al., 2025)

This challenge of metacognitive laziness is intensified by a key distinction in how people engage with generative AI—namely, their existing internal knowledge structures. Individuals with well-developed internal schemas—often those educated before AI became ubiquitous—can use these tools effectively. Their solid knowledge base allows them to evaluate AI output critically, refine prompts, integrate suggestions meaningfully, and detect inaccuracies. For these users, AI acts as a cognitive amplifier, extending their capabilities.

In contrast, learners still building foundational knowledge face a significant risk: mistaking AI fluency for their own. Without a robust internal framework for comparison, they may readily accept plausible-sounding output without realizing what's missing or incorrect. This bypasses the mental effort—retrieval, error detection, integration—that neuroscience shows is essential for forming lasting memory engrams and flexible schemas. The result is a false sense of understanding: the learner feels accomplished, but the underlying cognitive work hasn't been done.

This effect extends beyond the classroom. Lee and colleagues (2025) found that knowledge workers who placed high confidence in generative AI engaged in significantly less critical thinking, effectively offloading cognitive effort to the technology. The pattern is especially pronounced in areas requiring procedural skill development. In mathematics education, Bastani et al. (2024) reported that high school students who used GPT-4 during practice outperformed peers—until the final exam, when the AI was removed. Their performance dropped, revealing a lack of retained understanding. Similarly, Yang et al. (2025) found that students using ChatGPT for C++ programming experienced lower flow, reduced self-efficacy, and poorer learning outcomes compared to conventional learners. In both cases, students had used AI as a "crutch," bypassing the mental effort needed to transition from declarative to procedural memory. Without this shift, they failed to develop the intuitive grasp that emerges through repeated practice and error correction. The AI short-circuited the very cognitive labor that feels hard in the moment—but is essential for building the neural architecture of lasting understanding.

These studies highlight the complex relationship between AI tools and learning outcomes. The pattern is consistent across domains: when AI circumvents the brain's natural learning mechanisms—prediction errors, schema formation, and the crucial transition from declarative to procedural knowledge—learning suffers despite seemingly productive short-term outcomes. The technology creates an illusion of knowledge while bypassing the cognitive processes necessary for genuine learning.

Recent research on AI-generated educational tools reveals an ironic phenomenon: as these systems learn from existing educational materials, they tend to reproduce the very pedagogical biases we have identified as potentially harmful. Chen et al. (2025) found that AI lesson-planning tools predominantly promote minimally guided approaches with limited opportunities for knowledge acquisition and practice—precisely the approaches our neuroscience evidence suggests may impair robust schema formation. This technological reinforcement of problematic pedagogical trends creates a concerning feedback loop: AI systems learn from educational materials influenced by constructivist ideas, then generate new materials that further propagate these approaches, all while claiming to represent "contemporary educational values." This underscores the importance of critically examining not just how we use technology in education, but also the embedded pedagogical assumptions these tools carry forward.

**8.2 Finding Balance: Appropriate Offloading and Educational Design**

Cognitive offloading isn't inherently negative—it's a natural and often helpful strategy (we use calendars for appointments and calculators for arithmetic). Writing itself is a productive form of cognitive offloading. By externalizing ideas onto paper, we free working memory to handle more complex thoughts. But writing is also procedural; good writing involves internalizing effective patterns, structures, and vocabulary. Benjamin Franklin famously improved his writing by reconstructing admired passages from memory using brief notes as cues. This approach allowed him to offload ideas to paper while simultaneously building internal schemata for effective communication. Problems arise only when offloading replaces the initial learning process. Experts with solid schemata can safely offload some details, but novices who offload everything never internalize essential knowledge.

The shift toward cognitive offloading in education is what research calls "pathological altruism"—good intentions that backfire. Teachers who promoted calculators and "just look it up" approaches sincerely wanted to free students from what seemed like needless memorization. But this well-meaning approach missed how stored knowledge builds the mental foundation needed for advanced thinking. The result? An educational approach that, despite kind intentions, undermined the very cognitive foundations needed for the higher-level reasoning it aimed to develop. (Oakley, 2013)

The concept of pathological altruism helps explain why educational reforms with the best intentions can sometimes lead to negative outcomes. When we understand this dynamic, we can better navigate the balance between cognitive offloading and internalized knowledge.

**8.3 Technology that Enhances Rather than Replaces Learning**

Interestingly, not all technological interventions undermine learning. When designed with sound cognitive principles in mind, AI can actually enhance rather than diminish educational outcomes. A 2024 study in Ghana found that students using an AI-powered math tutor on WhatsApp for just one hour per week showed substantial improvement in math scores compared to a control group. The key difference? This intervention was deliberately designed to provide scaffolded practice and hints, not just answers. Rather than encouraging passive consumption of information, it prompted active engagement with the material. This suggests that technology can support learning when it complements rather than replaces the brain's natural learning mechanisms. (Henkel *et al.*, 2024)

Our message is not to reject technology or return to endless drills. Instead, research supports using technological tools in ways that complement human learning mechanisms, enhancing rather than inhibiting deep learning. The key is balance, guided by the "Eighty Five Percent Rule"—finding the sweet spot of challenge where students are pushed but not overwhelmed. Just as you can't build a house starting with the roof, the brain struggles to grasp advanced concepts without first mastering the basics. Calculators and AI tools that let students skip this foundation make higher-level learning inherently unstable. Likewise, the well-intentioned emphasis on "desirable difficulty" has sometimes pushed students beyond their optimal learning zone, leaving them frustrated and unable to form coherent mental frameworks. The goal should always be to preserve deep learning by ensuring students engage their own memory and reasoning at just the right level of challenge—not too easy, and not too difficult.

Based on the neuroscientific principles we've explored throughout this chapter, we propose the following evidence-based strategies for educators seeking to balance technological integration with cognitive development:

- Embrace desirable difficulty—within limits: Encourage learners to generate answers and grapple with problems before turning to help, but be mindful of maintaining that sweet spot of approximately 85% success. Struggling productively with challenges triggers the brain's natural learning enhancements, but excessive struggle can prevent the formation of effective neural manifolds. In classroom practice, this means carefully calibrating when to provide guidance—not immediately offering solutions, but also not leaving students floundering with tasks far beyond their current capabilities. Teachers can cultivate perseverance while still ensuring students experience enough success to build confidence and effective schemata. This approach supports what schema theorists call "tuning"—the gradual evolution of existing schemata to better represent the population of situations to which they apply, making them increasingly accurate and efficient.
- Assign foundational knowledge for memorization and practice: Rather than viewing factual knowledge as rote trivia, recognize it as the glue for higher-level thinking. This doesn't mean endless drilling without context, but it does mean certain basics (like math facts, vocabulary, scientific terms, historical dates, formulas) should be overlearned to the point of automaticity, helping that crucial transition from declarative to procedural memory we discussed earlier. Tools like spaced repetition software can combine technology with memory science to help students efficiently retain core knowledge. Far from being a waste of time, this internal library will enable students to think faster and more creatively later on.
- Use procedural training to build intuition: Allocate class time for practicing skills without external aids. For instance, mental math exercises, handwriting notes, reciting important passages or proofs from memory, and so on. Such practices, once considered old-fashioned, actually cultivate the procedural fluency that frees the mind for deeper insight. As an analogy, even though modern pilots have autopilot, they still rigorously train on manual flying – because that ingrains understanding and prepares them for when automation fails. Similarly, students who practice reasoning unaided develop a *feel* for it that no on-demand answer can impart.
- Intentionally integrate technology as a supplement, not a substitute: When using AI tutors or search tools, structure their use so that the student remains cognitively active. For example, an AI could be used to get hints or check work, rather than to produce the entire answer. Studies have shown that learning improves when AI is used under guidance (e.g., an AI that asks the student questions or encourages reflection, as opposed to an oracle that spoon-feeds answers). Teachers and software designers should aim for interactive engagement, where the AI can handle repetitive aspects or provide adaptive feedback, but the student must still do the reasoning or recall the knowledge. This keeps the learner in the driver's seat mentally.
- Promote internal knowledge structures: Help students build robust mental frameworks by ensuring connections happen inside their brains, not just on paper. While activities like concept mapping might appear helpful, they often create an illusion of understanding—students connect ideas externally without developing the neural manifolds that truly matter. As we explored earlier, the brain's procedural memory system and basal ganglia require direct internal processing to form durable cognitive structures. Instead, guide students to identify relationships between concepts through active questioning ("How does this principle relate to what we learned last week?") and guided reflection. When students use internet research, frame it as supplementary to their internal knowledge base, not a replacement. This approach builds on our understanding of prediction errors and schema development—information must be processed through the brain's own architecture to become truly functional knowledge that can trigger error detection and support intuitive understanding.

• Educate about metacognition and the illusion of knowledge: Help students recognize that *knowing where to find information is fundamentally different from truly knowing it*. Information that exists "out there" doesn't automatically translate to knowledge we can access and apply when needed. Guide students to reflect on their own learning processes so they can make more intentional choices about what to commit to memory. For example, a teacher might model: "Let's approach this problem using only what we know first. If we reach an impasse, we'll look up what we need, but then continue solving it on our own." This balanced approach treats external resources as supplements to internal knowledge—tools that support thinking rather than replace it.

**8.4 The Future of Learning: Integration Not Replacement**

In implementing these strategies, the goal is to cultivate learners who are both tech-savvy and deeply knowledgeable. The two are not mutually exclusive—they're mutually reinforcing. A student with strong internal memory and well-honed thinking skills will use technology *more effectively* than one who uses it as a crutch. They will know what to ask, how to evaluate responses, and how to integrate new information into existing knowledge structures. By contrast, a student trained to constantly offload cognitive processes faces an inevitable ceiling—passively consuming information without developing the neural architecture needed for true understanding, and faltering when external supports are unavailable.

Cognitively, the human brain remains both the ultimate bottleneck and the ultimate engine for learning. External devices expand our capabilities, but what happens in our neurons determines how far those capabilities go. An insight not formed is a memory not made. A skill not practiced is an intuition not developed. A schema not built is a problem not recognized. And a mind unchallenged is a talent unlived. The trends of the past decades serve as a caution: even as the world's knowledge sits readily accessible on every smartphone, we must ensure that knowledge also takes root *within the individual's mind*.

Looking ahead, the synergy of human cognition and machine capability will define successful education. We should strategically leverage AI and vast information stores to expand learning possibilities—personalizing instruction and creating immersive simulations—while preserving the core of what makes us intelligent**.** This essential core is our capacity to learn, remember, and reason independently of external support. It's the spark that produces genuine *insight*, the mental dexterity to solve problems within our own minds, and the wisdom to connect ideas spontaneously—cognitive processes that remain distinctly valuable even as technology increasingly mimics these abilities.

In conclusion, storing key information in human memory is not a quaint educational ideal; it is a pillar of cognitive function. The neuroscience we explored validates age-old common sense: *we remember what we wrestle with, we excel at what we practice, and we understand what we internalize.*

Fully exploring motivation, curiosity, and enthusiasm would substantially expand this chapter. Yet it is worth briefly noting the paradox highlighted by memory researchers Wang and Morris (2009): "we rapidly remember what interests us, but what interests us takes time to develop." In other words, genuine interest—and thus deep learning—often emerges gradually, as we become proficient, rather than preceding that proficiency. Ironically, in today's rush to make learning instantly appealing through games, rewards, or engaging demonstrations, we can overlook this crucial insight. Students initially attracted to STEM fields by exciting experiments or playful activities in high school may struggle when confronted with challenging coursework in college—the "math-science death march" described by David Goldberg (Drew, 2011). Lasting interest, the kind essential for expertise, is typically nurtured slowly, sustained by the satisfaction and intrinsic reward of growing competence.

Reflecting historically, despite their significant contributions, influential educational figures have sometimes obscured our understanding of how learning truly unfolds. Piaget, Vygotsky, Dewey, Bloom, Montessori, Bruner, and Skinner, among others, proposed groundbreaking ideas that captured imaginations and reshaped classrooms worldwide. Yet, as neuroscience reveals, many core assumptions of these thinkers—such as the effectiveness of unguided discovery learning, procedural-only approaches, or undervaluing explicit memorization—have hindered more effective educational practices. Appreciating both the insights and the limitations of these influential figures can guide educators toward more evidence-based, neurologically informed teaching methods, ultimately benefiting learners.

Rather than viewing memory as obsolete in the AI era, we should treat it as our personal knowledge bank. This empowers us to use AI wisely. By thoughtfully balancing what our minds and machines each do best, we ensure that external innovations enhance rather than diminish our intelligence. Ideally, future learners will confidently say: "I use the internet, but I don't have to look everything up, because I've learned and remembered what matters." Such deep, resilient knowledge will be essential in navigating a world of endless information—and protecting our minds from cognitive decline amid constant technological distractions.

50Wickelgren, I. (2025) How 'event scripts' structure your personal memories, *Quanta Magazine*. Available at: https://www.quantamagazine.org/how-event-scripts-structure-our-personal-memories-20250221/

Wilson, R. C., Shenhav, A., Straccia, M. and Cohen, J. D. (2019) The eighty five percent rule for optimal learning, *Nat Commun,* 10(1), pp. 4646. 10.1038/s41467-019-12552-4

Wimmer, G. E. and Büchel, C. (2021) Reactivation of single-episode pain patterns in the hippocampus and decision making, *The Journal of Neuroscience*, 41(37), pp. 7894-7908. 10.1523/jneurosci.1350-20.2021

Wojcik, M. J., Li, A. X., Wasmuht, D., Stroud, J. P., Stokes, M. G., Myers, N. E. and Hunt, L. T. (2025) Working memory shapes neural geometry in human EEG over learning, *bioRxiv*, pp. 2025.01. 21.634110

Yang, Y., Huang, Z., Yang, Y., Fan, M. and Yin, D. (2025) Time-dependent consolidation mechanisms of durable memory in spaced learning, *Commun Biol*, 8(1), pp. 535. 10.1038/s42003-025-07964-6

Yang, T.-C., Hsu, Y.-C., and Wu, J.-Y. (2025) The effectiveness of ChatGPT in assisting high school students in programming learning: evidence from a quasi-experimental research, *Interactive Learning Environments.* doi.org/10.1080/10494820.2025.2450659

Yang, W., Sun, C., Huszár, R., Hainmueller, T., Kiselev, K. and Buzsáki, G. (2024) Selection of experience for memory by hippocampal sharp wave ripples, *Science*, 383(6690), pp. 1478-1483

Ye, L., Ba, L. and Yan, D. (2025) A study of dynamic functional connectivity changes in flight trainees based on a triple network model, *Scientific Reports,* 15(1). 10.1038/s41598-025-89023-y